\documentclass[
aps, prd, reprint, 
10pt, notitlepage, a4paper,
floats, floatfix,
amsmath, amssymb, amsfonts, 
superscriptaddress,
showpacs, showkeys,
nofootinbib,
longbibliography,
]{revtex4-1}

\usepackage[utf8]{inputenc}

\usepackage[usenames,dvipsnames]{xcolor}

\xdefinecolor{mylinkcolor}{rgb}{0,0,0.5}
\usepackage[
	bookmarksnumbered, bookmarksopen, bookmarksopenlevel=2,
	breaklinks=true,
	colorlinks=true, filecolor=mylinkcolor, citecolor=mylinkcolor,
	linkcolor=mylinkcolor, urlcolor=mylinkcolor, menucolor=mylinkcolor,
]{hyperref}

\allowdisplaybreaks

\usepackage{feynmp-auto}
\usepackage{feynmp}
\def\vct#1{\mathbf{#1}}

\def\nl{\\ & \quad}
\def\nnl{\nonumber \\ & \quad}

\DeclareMathOperator{\Order}{\mathcal{O}}

\newcommand\be{\begin{equation}}
\newcommand\ee{\end{equation}}
\newcommand\bea{\begin{align}}
\newcommand\eea{\end{align}}

\newcommand{\bal}[1]{\setlength{\jot}{1em}\begin{align}\begin{aligned}#1\end{aligned}\end{align}}

\newcommand{\ttau}{\tilde\tau}

\begin{document}

\title{Effective action of dilaton gravity as the classical double copy of Yang-Mills theory}

 \author{Jan Plefka}
 \email{jan.plefka@physik.hu-berlin.de}
 \affiliation{Institut f\"ur Physik und IRIS Adlershof, Humboldt-Universit\"at zu Berlin, 
   Zum Gro{\ss}en Windkanal 6, D-12489 Berlin, Germany}
 \author{Jan Steinhoff}
 \email{jan.steinhoff@aei.mpg.de}
 \affiliation{Max Planck Institute for Gravitational Physics (Albert Einstein Institute), Am M\"uhlenberg 1, Potsdam D-14476, Germany}
 \author{Wadim Wormsbecher}
 \email{wadim.wormsbecher@physik.hu-berlin.de}
 \affiliation{Institut f\"ur Physik und IRIS Adlershof, Humboldt-Universit\"at zu Berlin, 
  Zum Gro{\ss}en Windkanal 6, D-12489 Berlin, Germany}

\preprint{HU-EP-18/24}

\begin{abstract}
  We compute the classical effective action of color charges moving along 
  worldlines by integrating out the Yang-Mills gauge field to next-to-leading order in the coupling.
  An adapted version of the Bern-Carrasco-Johansson (BCJ) double-copy construction
  known from quantum scattering amplitudes is then
  applied to the Feynman integrands, yielding the prediction for the classical effective action
   of point masses in dilaton gravity.  
 We check the validity of the result by independently constructing the effective
 action in dilaton gravity employing field redefinitions and gauge choices that greatly simplify the
 perturbative construction. Complete agreement is found  at next-to-leading order. Finally, upon performing the post-Newtonian
 expansion of our result, we find agreement with the corresponding action of scalar-tensor
 theories known from the literature. Our results represent a proof of concept for the
 classical double-copy construction of the gravitational effective action and provides
 another application of a BCJ-like double copy beyond scattering amplitudes.
\end{abstract}

\maketitle

\section{Introduction}

There is a growing body of evidence for a fascinating perturbative duality between Yang-Mills theory
and  quantum gravity known as the double-copy construction or color-kinematics duality
due to Bern, Carrasco and Johansson (BCJ)
\cite{Bern:2008qj, Bern:2010yg, Bern:2010ue}. 
It provides a concrete prescription for transforming scattering amplitudes in non-Abelian
gauge theories into scattering amplitudes in gravitational theories
 upon replacing the non-Abelian color degrees of freedom 
by kinematical ones. 
In the simplest scenario pure Yang-Mills theory double copies to dilaton gravity
coupled to an axion also termed as $\mathcal{N}=0$ supergravity (the massless sector
of bosonic strings).
This relation was proven for tree-level amplitudes \cite{Bern:2010yg}, where the 
double copy is equivalent to the Kawai-Lewellen-Tye relations of string theory \cite{Kawai:1985xq},
a precursor of the BCJ duality. 
It is also  manifest in the Cachazo-He-Yuan formulation of gluon and graviton tree-level amplitudes \cite{Cachazo:2014xea}.
While remaining conjectural at the loop level, the double-copy procedure is extremely efficient in generating integrands for gravitational theories
at high-perturbative orders in theories with and without supersymmetry
\cite{Bern:2011rj,Bern:2012uf,Carrasco:2012ca,Bern:2013yya,Chiodaroli:2014xia,Chiodaroli:2015wal,Mogull:2015adi,He:2015wgf,Johansson:2014zca} and
as such has passed many nontrivial checks. The present record being at the four point five-loop level
for maximal supergravity \cite{Bern:2018jmv}. Moreover, an elaborate web of theories connected via
color-kinematics duality exists that includes matter couplings and various numbers of supersymmetries \cite{Anastasiou:2016csv,Anastasiou:2017nsz,Cheung:2017ems}. All these results point towards the double copy being a generic property of gravity alluding at a hidden kinematical algebra which has resisted discovery 
so far.

In light of these findings the natural question arises, whether the double copy generalizes beyond the
realm of scattering amplitudes. In particular, does it also play a role in classical general
relativity? Here a number of encouraging results have been obtained: Schwarzschild, Kerr
and Taub-NUT spacetimes were shown to be double copies of classical gauge-theory solutions
\cite{Monteiro:2014cda,Luna:2015paa}, were further extended to certain classes
of perturbative spacetimes \cite{Luna:2016hge}, and a classical double-copy description of the
gravitational radiation from accelerating black holes was developed in Ref.~\cite{Luna:2016due}.
Following up the last direction, a very interesting setting is that of
gravitational radiation produced by binary sources, not the least due to the spectacular
observations of gravitational waves and the need for high precision predictions to generate
waveform templates \cite{Abbott:2016blz}. 
Pioneering work  was done in 
\cite{Goldberger:2016iau,Goldberger:2017vcg, Goldberger:2017ogt}
where perturbative classical solutions for binary color-charged point particles coupled via Yang-Mills
theory \cite{Balachandran:1976ya} were shown to double copy to their counterparts in 
dilaton-gravity radiation  (see also \cite{Goldberger:2017frp,Chester:2017vcz}). In these
works certain color-kinematics replacement rules were employed which appear somewhat distinct
to the color factor/numerator replacement rules familiar from scattering amplitudes.
Also the question of a double-copy-respecting representation of the perturbative gauge-theory solutions remained open. This situation was clarified very recently in the  work of Shen \cite{Shen:2018ebu}, which pushed the perturbative approach of solving the equations of motions
via double copy to the next-to-leading order in the coupling constant
expansion. In this process a color-kinematics duality representation of the
Yang-Mills-radiation solution was found, which parallels the color-kinematics duality
rules of BCJ \cite{Bern:2008qj} and replaces the nonstandard double-copy rules of \cite{Goldberger:2016iau}.
An alternative route for finding the graviton radiation of binaries via the double copy was 
taken in \cite{Luna:2017dtq}, where a tree scattering amplitude in gauge theory coupled to scalar fields was double copied to establish the leading order gravitational radiation emitted 
from the scattering of two black holes (modeled by two massive scalar fields).

In our present work, we generalize these approaches to the double copy of gravitationally interacting binaries by ascending from the level of equations of motion 
to the classical effective action. This approach makes direct contact to the 
post-Minkowskian (weak-field) and post-Newtonian (weak-field and slow-motion)
expansions of the gravitational potential for which 
high-order results exist in the literature, namely at the fourth post-Newtonian order (four loop)
for nonspinning bodies: using a canonical formalism of general relativity
\cite{Damour:2014jta}, a Fokker Lagrangian \cite{Bernard:2016wrg,Marchand:2017pir}, and
partial results within an effective field theory formalism \cite{Foffa:2016rgu}.
The latter makes use of Feynman diagrammatic methods and sophisticated tools for effective
field theories in the context of classical gravity \cite{Goldberger:2004jt}.
The question how the classical gravitational potential may be extracted 
from the quantum scattering amplitude of massive scalars has in fact a long history
starting in the 1970s
\cite{Iwasaki:1971vb,Duff:1973zz,Holstein:2004dn}. Recent works have updated this by employing modern
unitarity methods of quantum field theory \cite{Neill:2013wsa,Bjerrum-Bohr:2013bxa,Bjerrum-Bohr:2018xdl}.
Damour recently proposed an alternative approach for converting scattering angles and amplitudes
to the effective one-body Hamiltonian  \cite{Damour:2016gwp,Damour:2017zjx}.
Hence, in order to perform the construction of the interaction potential using the double copy,
one could study the scattering amplitude of a massive scalar field coupled to either Yang-Mills or
dilaton-gravity theory and relate to a potential using one of the mentioned works.
Representing the gauge-theory result in a color-kinematics respecting fashion should then
yield the effective gravitational potential via a suitably defined double copy.
Here we follow a more direct approach, where the classical interaction potential
is the direct outcome of a path integral. For this purpose, we start out with color
charges moving along a worldline as in \cite{Goldberger:2016iau,Goldberger:2017vcg,
Goldberger:2017ogt, Shen:2018ebu}. This may be viewed as a classical limit of the
massive scalar field.

Our paper is organized as follows:
In Sec.~II we give an overview of the structure of a theory of classical color-charged massive point particles moving along worldlines, interacting via a Yang-Mills theory, in the first-order formalism.
In Sec.~III three we propose a double-copy prescription for the classical effective action. We compute the effective action of the theory introduced in Sec.~II up to next-to-leading order in the gluon coupling. Along the way we introduce the notion of a trivalent representation for graphs involving worldlines and at the end of the section we perform the double copy.
In Sec.~IV we repeat the computation of the effective action in the weak-field/post-Minkowskian approximation for massive point particles interacting in  dilaton-gravity theory and we compare the resulting expression to the double-copy result.
In Sec.~V we reexpand our result for the post-Minkowskian dilaton-gravity effective action in velocities, thus yielding the post-Newtonian approximation. After solving the Feynman integrals, we compare the resulting expression to known results in the literature.
Our conclusions are presented in Sec.~VI. Finally, our conventions, the Feynman rules and
a discussion of self-interactions can be found in the appendices.

\section{Yang-Mills interaction of color-charged massive particles}
We consider the worldline action of a massive, color-charged (non-Abelian) point particle
coupled to the Yang-Mills gauge field $A_{\mu}^{a}$; see appendix \ref{conventions}
for notation and conventions. The action for the classical colored point charge (pc)
is given by
\cite{Balachandran:1976ya,Goldberger:2016iau, Goldberger:2017vcg}
\begin{align}
  S_\text{pc} &= - \int d\tau L_\text{pc} = - \int d\tau \left( m \sqrt{u^2} - \psi^\dagger i u^\mu D_\mu \psi \right) \\
  &= - \int d\tau \left( m \sqrt{u^2} - i \psi^\dagger \dot \psi - g u^\mu A_\mu^a c^a \right) \, ,
\end{align}
which is invariant under reparametrization of $\tau$.
Here $u^{\mu}= \dot x^{\mu}$ is the 4-velocity of the particle along the
worldline $x^{\mu}(\tau)$ and
$\psi(\tau)$ an associated fundamental vector carrying the color degrees
of freedom of the particle -- often called the color wave function. Moreover
\be
c^{a}:= \psi^{\dagger}T^{a}\psi\, ,
\ee
is the color charge carried by the particle.
In this form, there is only a single gluon coupling to the scalar worldline.
As the gravity counterpart also has higher graviton couplings to the worldline 
of a massive particle, this poses an immediate obstacle to a double-copy relation.
Also a massive scalar field, for which a double copy of scattering amplitudes
exist \cite{Luna:2017dtq}, has linear \emph{and quadratic} couplings to gluons.
However, in a first-order formalism this situation is remedied. Defining canonical momenta
\begin{equation}
  p_{\mu} = \frac{\partial L_\text{pc}}{\partial u^\mu}
             = m \frac{u_{\mu}}{\sqrt{u^2}} - g A_\mu^a c^a , \label{pcan}
  \;\; \Rightarrow \;\; m^2 = (p + g A c)^2 , 
\end{equation}
we find the Hamiltonian and the action in the first-order formalism
\begin{align}
  H_\text{pc} &= H_{\text{can}} + \lambda \left[ (p + g A c)^2 - m^2 \right] \, , \nonumber \\
   H_{\text{can}} &= p_{\mu} u^\mu - L_{pc} \stackrel{\eqref{pcan}}{=} i \psi^\dagger \dot \psi , \\
  S_\text{pc} &= - \int d\tau ( p_{\mu} u^\mu - H_\text{pc} ) \nonumber \\
  &= - \int d\tau \Big( p_{\mu} u^\mu - i \psi^\dagger \dot \psi
    - \lambda \big[ p^2 + 2 g p_{\mu} A^\mu_a c^a \nnl
    + g^2 A_\mu^b c^b A^\mu_a c^a - m^2 \big] \Big) .
\end{align}
Here, $H_{\text{can}}$ is the ``canonical'' Hamiltonian  and
$H_\text{pc}$ is the ``Dirac'' Hamiltonian where the mass-shell constraint was added using a
Lagrange multiplier $\lambda(\tau)$. Now we have both a single and a double gluon coupling to the
worldline, similar to a massive scalar field.
Note that $\lambda(\tau) d\tau$ is reparametrization invariant.

We now consider Yang-Mills gauge theory coupled to a 
set of two scalar particles in the worldline description. Notationally we separate the worldlines by placing a tilde on all variables associated to one of the worldlines. Therefore the full action for classical Yang-Mills (cYM) reads
\be
S_\text{cYM}=S_{\text{YM}}+ S_{\text{gf}}+ S_\text{pc}+ \tilde S_\text{pc} \, ,
\ee
where a gauge fixing part $S_{\text{gf}}$ was added.
With this, the gluon propagator reads
\bal{
\langle A_{\mu}^{a}(x)\, A_{\nu}^{b}(y)\rangle_{0}\, =\,  \frac{\hbar}{i}\, \eta_{\mu\nu}\, \delta_{ab}\, D(x-y)~,
}
where $D(x-y)$ is defined via
\bal{
\Box D(x-y) = -\delta(x-y)~.
}

\section{Effective action of colored massive particles and its double copy}

\subsection{Principle of the double-copy construction}
\label{section:dcp}
The double-copy scheme to be applied here is to first 
integrate out the gauge field $A^{a}_{\mu}$ in order to obtain the
effective action $S_\text{eff,YM}$ for two colored, massive particles with Yang-Mills
interactions, i.e.
\begin{equation}
  e^{\frac{i}{\hbar} S_\text{eff,YM}} =  e^{\frac{i}{\hbar} S_\text{pc,free}} \mathcal{M}_{\text{YM}}
  = \text{const} \int \mathcal{D} A \, e^{\frac{i}{\hbar} S_\text{cYM}} \, .
\end{equation}
The (divergent) normalization constant on the right-hand side is chosen such that $\mathcal{M}_\text{YM} = 1$ for $g \rightarrow 0$ and
$S_\text{pc,free}$ is given by $S_\text{pc} + \tilde{S}_\text{pc}$ for $g \rightarrow 0$.
The perturbative expansion in the coupling $g$ of $\mathcal{M}_{\text{YM}}$ can always be brought into the schematic form
\begin{equation}\label{eq:DCstart}
  \mathcal{M}_{\text{YM}} = \sum_{n=0}^{\infty} (2g)^{2n} 
  \sum_{I\in\Gamma_n} \left( \frac{i}{\hbar} \right)^{x_{I}} 
  \int \prod_{i_{I}} d\hat\tau_{i_{I}} \int d^{4l_{I}}x
  \frac{C_I N_I}{S_I D_I} ,
\end{equation}
where $\Gamma_n$ is the set of N$^{n-1}$LO trivalent graphs with two external worldlines
of the colored point particles, $C_I$ denotes the color factor associated with graph $I$, i.e.~functions of the $c(\tau)$'s,
$N_{I}$ the associated kinematic numerator factors and $D_I$ the spacetime propagators appearing in the graph $I$. The $x_I$ is defined as the number of vertices (bulk and worldline) minus the number of propagators in the trivalent diagram and $l_I$ is the number of bulk vertices. Note that the integral measure along the worldline $d\tau$ always comes with
a corresponding Lagrange multiplier $\lambda(\tau)$ in order to guarantee reparametrization
invariance. We collect this into a new
density notation via
\be
d\hat\tau_{i} := d\tau_{i} \, \lambda(\tau_{i})\,.
\ee 
The $S_{I}$ finally, denote the symmetry factor of the graph $I$ in the trivalent representation, i.e.~the graph has to be drawn in such a way that it reproduces the color structure $C_I$. We will explain our notion of a trivalent representation more precisely in the next section.

The double-copy prescription that we are adopting here then amounts to representing the
exponential of the effective action of two massive particles coupled to a weak-field (post-Minkowskian) expansion of dilaton gravity (dg) as
\begin{equation}
  \mathcal{M}_{\text{dg}} = \sum_{n=0}^{\infty} (i\kappa)^{2n} 
  \sum_{I\in\Gamma_n}  \left( \frac{i}{\hbar} \right)^{x_{I}} 
 \int \prod_{i_{I}} d\hat\tau_{i_{I}} \int d^{4l_{I}}x
 \frac{N_I N_I}{S_I D_I} ,\label{eq:DCprescription}
\end{equation}
where $\kappa$ is the gravitational coupling constant.
After that, the effective action in dilaton gravity follows from $e^{\frac{i}{\hbar} S_\text{eff,dg}} = e^{\frac{i}{\hbar} S_\text{pc,free}} \mathcal{M}_{\text{dg}}$.
Only classical terms ($\hbar \rightarrow 0$) are retained in $S_\text{eff,dg}$ in our
considerations. At higher order,
$C_I$ and $N_I$ are required to fulfill the usual BCJ color-kinematics duality (Jacobi relations). We conclude this section by stressing that our double-copy prescription is proposed at the level of $\mathcal M$ and not at the level of the effective action.  

\subsection{The trivalent representation of the effective action of colored particles to next-to-leading order\label{YMTECH}}

\begin{fmffile}{EffectivePotential}
We now compute the effective action for two color-charged massive point particles moving
on their worldlines to next-to-leading order, $\mathcal O(g^4)$. Furthermore, we will clarify what we mean by a trivalent representation of a diagram involving worldlines. We will use the short-hand notation
\begin{align}
  c_i &:= c(\tau_i) , \quad p_i := p(\tau_i) , \nonumber \\
  D_{ij} &:= D(x(\tau_i)-x(\tau_j)) , \quad d\hat\tau_{1...n} := \prod_{i=1}^{n}\,d\tau_i \,\lambda(\tau_i)\,.
\end{align}
Additionally, we will use the tilde notation for the ``right" worldline in our diagrams.

At leading order we have the three graphs
\begin{align}
\raisebox{-0.6cm}{\begin{fmfgraph*}(50,40)
	\fmfpen{thin}
	\fmfleft{a1,a2}
	\fmfright{b1,b2}
	\fmf{vanilla}{a1,v1,a2}
	\fmf{vanilla}{b1,v2,b2}
	\fmffreeze
	\fmf{photon}{v1,v2}
	\fmfdot{v1,v2}
	\fmfv{label=$\tau_1$}{v1}
	\fmfv{label=$\tilde\tau_2$}{v2}
\end{fmfgraph*}}\qquad 
&= 4  g^{2}\frac{i}{\hbar}\,\int d\hat\tau_{1\tilde2} (c_{1}\cdot \tilde c_{2})\,(p_{1}\cdot \tilde p_{2})\, 
\ D_{1\tilde2}\, ,
\\\nonumber\\
\raisebox{-0.6cm}{
\begin{fmfgraph*}(50,40)
	\fmfpen{thin}
	\fmfright{a1,a2}
	\fmfleft{b1,b2}
	\fmf{vanilla}{a1,v1,a2}
	\fmf{vanilla}{b1,v2,v5,b2}
	\fmffreeze
	\fmf{photon,right=1}{v2,v5}
	\fmfv{label=$\tau_1$}{v5}
	\fmfv{label=$\tau_2$}{v2}
	\fmfdot{v2,v5}
\end{fmfgraph*}} \qquad
&=  2  g^{2}\frac{i}{\hbar}\,\int d\hat\tau_{12}
(c_{1}\cdot c_{2})\,(p_{1}\cdot p_{2})\,  D_{12}\, ,
\\\nonumber\\
\raisebox{-0.6cm}{
\begin{fmfgraph*}(50,40)
	\fmfpen{thin}
	\fmfleft{b1,b2}
	\fmfright{a1,a2}
	\fmf{vanilla}{a1,v1,a2}
	\fmf{vanilla}{b2,v2,b1}
	\fmffreeze
	\fmf{phantom}{v1,z1,z2,z3,z4,v2}
	\fmf{photon,left=1}{v2,z1,v2}
	\fmfdot{v2}\fmfv{label=$\tau_{1}$}{v2}
\end{fmfgraph*}} \qquad
&\propto \int d\tau_{1} D_{11} = 0 \quad \text{in dim.~reg.}
\end{align}
Since the last term vanishes in dimensional regularization, we neglect any graph involving a bubble, as shown above, or any other scaleless integral. Note that we do not mention the mirrored counterparts to every graph with an uneven number of untilded and tilded variables. They can be trivially obtained by replacing untilded and tilded and it is understood that we add them to our final results.

At the next-to-leading order we encounter the graphs
\begin{widetext}
\begin{align}
\raisebox{-0.6cm}{
\begin{fmfgraph*}(50,40)
	\fmfpen{thin}
	\fmfleft{a1,a2}
	\fmfright{b1,b2}
	\fmf{vanilla}{a1,v1,v3,a2}
	\fmf{vanilla}{b1,v2,v4,b2}
	\fmffreeze
	\fmf{photon}{v1,v2}
	\fmf{photon}{v3,v4}
	\fmfdot{v1,v2,v3,v4}
	\fmfv{label=$\tau_1$}{v1}
	\fmfv{label=$\tilde\tau_2$}{v2}
	\fmfv{label=$\tau_3$}{v3}
	\fmfv{label=$\tilde\tau_4$}{v4}
\end{fmfgraph*}}\quad &=
8\,g^{4}\left(\frac{i}{\hbar}\right)^2\,\int d\hat\tau_{1\tilde 2 3 \tilde 4}\,
(c_{1}\cdot \tilde c_{2})\,(c_{3}\cdot \tilde c_{4}) \, (p_{1}\cdot \tilde p_{2})\,(p_{3}\cdot \tilde p_{4})\, D_{1\tilde2}\,D_{3\tilde 4}
\, ,
\\\nonumber\\
\qquad
\raisebox{-0.6cm}{\begin{fmfgraph*}(50,40)
	\fmfpen{thin}
	\fmfleft{a1,a2}
	\fmfright{b1,b2}
	\fmf{vanilla}{a1,v1,a2}
	\fmf{vanilla}{b1,v2,v4,b2}
	\fmffreeze
	\fmf{photon}{v1,v2}
	\fmf{photon}{v1,v4}
	\fmfdot{v1,v2,v4}
	\fmfv{label=$\tau_1$}{v1}
	\fmfv{label=$\tilde\tau_2$}{v2}
	\fmfv{label=$\tilde\tau_4$}{v4}
\end{fmfgraph*}}\quad
&= 4g^{4}\, \frac{i}{\hbar}\, \int d\hat\tau_{1\tilde 2\tilde 4}\, (c_{1}\cdot \tilde c_{2})\,(c_{1}\cdot \tilde c_{4})\,(\tilde p_{2}\cdot \tilde p_{4})\,  D_{1\tilde2}\, D_{1\tilde4}\nonumber
\\
&=4g^{4}\, \frac{i}{\hbar}\, \int d\hat\tau_{1\tilde 2\tilde 4 3 }\, (c_{1}\cdot \tilde c_{2})\,(c_{3}\cdot \tilde c_{4})\,(\tilde p_{2}\cdot \tilde p_{4})\,  D_{1\tilde2}\, D_{3\tilde4}\,\frac{\delta(\tau_3 - \tau_1)}{\lambda_3}\, ,
\end{align}
where we used a delta function in the last step in order to introduce a dummy $\tau_3$ integration. This can be understood as pulling apart the gluon-gluon-worldline vertex into two gluon-worldline vertices, i.e.~arriving at our definition of a trivalent representation necessary for our proposed double-copy prescription \eqref{eq:DCprescription}.

The next graph is
\begin{align}
\raisebox{-0.6cm}{
\begin{fmfgraph*}(50,40)
	\fmfpen{thin}
	\fmfleft{a1,a2}
	\fmfright{b1,b2}
	\fmf{vanilla}{a1,v1,a2}
	\fmf{vanilla}{b1,v2,v3,b2}
	\fmffreeze
	\fmf{photon,tension=1.7}{v1,x1}
	\fmf{photon}{x1,v2}
	\fmf{photon}{x1,v3}
	\fmfdot{v1,v2,v3}
	\fmfv{decor.shape=circle,decor.filled=empty,decor.size=0.12w}{x1}
	\fmfv{label=$\tau_1$}{v1}
	\fmfv{label=$\tilde \tau_2$}{v2}
	\fmfv{label=$\tilde \tau_3$}{v3}
	\fmfv{label=$x$,l.a=90}{x1}
\end{fmfgraph*}}\quad
=\,
- 4\,g^{4}\, \frac{i}{\hbar} \int d\hat\tau_{1\tilde 2\tilde 3}\,  f^{abc}\, c_{1}^{a}\tilde c_{2}^{b}\tilde c_{3}^{c}\,
V^{\mu\nu\rho}_{1\tilde2\tilde3}\,p_{1\mu}\,\tilde p_{2\nu}\,\tilde p_{3\rho}\, G_{1\tilde2\tilde3}\, ,\label{eq:ThreeGluonVert}
\end{align}
where $V_{123}^{\mu_1\mu_2\mu_3}$ is the color independent part of the Yang-Mills three gluon vertex
\bal{\label{defV}
	V^{\mu_1\mu_2\mu_3}_{123} =	\eta^{\mu_{1}\mu_{2}}\, (
	\partial_{1}^{\mu_{3}}-\partial_{2}^{\mu_{3}}) + 
	\text{cyclic}~, 
}
and $G_{1\tilde2\tilde3}= \int d^{4}x D_{1x}D_{\tilde2x}D_{\tilde3x}$.
We proceed in an analogous fashion with
\begin{align}
\raisebox{-0.6cm}{
\begin{fmfgraph*}(50,40)
	\fmfpen{thin}
	\fmfleft{a1,a2}
	\fmfright{b1,b2}
	\fmf{vanilla}{a1,v1,a2}
	\fmf{vanilla}{b1,v2,v4,b2}
	\fmffreeze
	\fmf{photon}{v1,v2}
	\fmf{photon,right=1}{v2,v4}
	\fmfdot{v1,v2,v4}
	\fmfv{label=$\tau_1$}{v1}
	\fmfv{label=$\tilde\tau_2$,l.a=-25}{v2}
	\fmfv{label=$\tilde\tau_3$}{v4}
\end{fmfgraph*}}\quad
&=  8\,  g^{4} \frac{i}{\hbar}\, \int d\hat\tau_{1\tilde 2 \tilde 3}\,(c_{1}\cdot \tilde c_{2})\,(\tilde c_{2}\cdot \tilde c_{3})\,
(p_{1}\cdot \tilde p_{3})\, D_{1\tilde2}\, D_{\tilde2\tilde3}\nonumber \\
&=8\,  g^{4} \frac{i}{\hbar}\, \int d\hat\tau_{1\tilde 2 \tilde 3\tilde 4}\,(c_{1}\cdot \tilde c_{2})\,(\tilde c_{4}\cdot \tilde c_{3})\,
(p_{1}\cdot \tilde p_{3})\, D_{1\tilde2}\, D_{\tilde4\tilde3}\frac{\delta(\ttau_4-\ttau_2)}{\tilde\lambda_4}~,\\\nonumber\\
\raisebox{-0.6cm}{
\begin{fmfgraph*}(50,40)
	\fmfpen{thin}
	\fmfleft{a1,a2}
	\fmfright{b1,b2}
	\fmf{vanilla}{a1,v1,a2}
	\fmf{vanilla}{b1,v2,vn,v4,b2}
	\fmffreeze
	\fmf{photon}{v1,vn}
	\fmf{photon,right=1}{v2,v4}
	\fmfdot{v1,v2,v4,vn}
	\fmfv{label=$\tau_1$}{v1}
	\fmfv{label=$\tilde\tau_4$}{v2}
	\fmfv{label=$\tilde\tau_2$,l.a=135}{vn}
	\fmfv{label=$\tilde\tau_3$}{v4}
\end{fmfgraph*}}\quad
&= 
8\, g^{4}\left (\frac{i}{\hbar}\right )^{2}\, 
\int d\hat \tau_{1\tilde 2\tilde 3\tilde 4}\, 
(c_{1}\cdot \tilde c_{2})\,(\tilde c_{3}\cdot \tilde c_{4})\,
(p_{1}\cdot \tilde p_{2})\,(\tilde p_{3}\cdot \tilde p_{4})\, D_{1\tilde2}\, D_{\tilde3\tilde4}\, ,
\\\nonumber\\
\raisebox{-0.6cm}{
\begin{fmfgraph*}(50,40)
	\fmfpen{thin}
	\fmfleft{a1,a2}
	\fmfright{b1,b2}
	\fmf{vanilla}{b1,b2}
	\fmf{vanilla}{a1,v2,a,v,k,l,j,v4,r,w,v5,g,h,o,p,u,v6,a2}
	\fmf{photon,right=1,tension=0}{v2,v4}
	\fmf{photon,right=1,tension=0}{v5,v6}
	\fmfdot{v2,v4,v5,v6}
	\fmfv{label=$\tau_1$}{v4}
	\fmfv{label=$\tau_4$}{v5}
	\fmfv{label=$\tau_2$,l.a=180}{v2}
	\fmfv{label=$\tau_3$,l.a=180}{v6}
\end{fmfgraph*}
} ~~ &= 2\, g^{4}\, \left (\frac{i}{\hbar}\right )^{2} \int d\hat\tau_{1234} \, 
(c_{1}\cdot  c_{2})\,( c_{3}\cdot  c_{4})\,
( p_1 \cdot  p_2) ( p_3 \cdot  p_4)\,  D_{12} D_{43} ~, 
\\\nonumber\\
\raisebox{-0.6cm}{
\begin{fmfgraph*}(50,40)
	\fmfpen{thin}
	\fmfleft{a1,a2}
	\fmfright{b1,b2}
	\fmf{vanilla}{b1,v1,b2}
	\fmf{vanilla}{a1,v2,g,h,j,v4,a,s,d,v5,a2}
	\fmffreeze
	\fmf{photon,right=1,tension=0}{v2,v4}
	\fmf{photon,right=1,tension=0}{v4,v5}
	\fmfdot{v2,v4,v5}
	\fmfv{label=$\tau_1$}{v4}
	\fmfv{label=$\tau_2$,l.a=180}{v2}
	\fmfv{label=$\tau_3$,l.a=180}{v5}
\end{fmfgraph*}
} ~~ &= 4\, g^{4}\, \frac{i}{\hbar}  \int d\hat\tau_{123} \, 
( c_{1}\cdot  c_{2})\,( c_{1}\cdot  c_{3})\,
( p_2 \cdot  p_3) \,  D_{12} D_{13} \nonumber\\
&= 4\, g^{4}\, \frac{i}{\hbar}  \int d\hat\tau_{1234} \, 
( c_{1}\cdot  c_{2})\,( c_{4}\cdot  c_{3})\,
( p_2 \cdot  p_3) \,  D_{12} D_{43}\,\frac{\delta(\tau_4 - \tau_1)}{\lambda_4}
~ . 
\end{align} 
\end{widetext}
We also have graphs that are not mirrored counterparts of above mentioned graphs, i.e. 
\begin{align}
\raisebox{-0.6cm}{
	\begin{fmfgraph}(50,40)
	\fmfpen{thin}
	\fmfleft{a1,a2}
	\fmfright{b1,b2}
	\fmf{vanilla}{b1,v2,v4,b2}
	\fmf{vanilla}{a1,v5,v6,a2}
	\fmf{photon,left=1,tension=0}{v2,v4}
	\fmf{photon,right=1,tension=0}{v5,v6}
	\fmfdot{v2,v4,v5,v6}
	\fmfv{label=$\tau_1$}{v4}
	\fmfv{label=$\tau_4$}{v5}
	\fmfv{label=$\tau_2$,l.a=180}{v2}
	\fmfv{label=$\tau_3$,l.a=180}{v6}
	\end{fmfgraph}
}\qquad,\qquad 
\raisebox{-0.6cm}{
	\begin{fmfgraph}(50,40)
	\fmfpen{thin}
	\fmfleft{a1,h1,a2}
	\fmfright{b1,b2}
	\fmf{vanilla}{a1,a2}
	\fmf{vanilla}{b1,v2,v1,v3,b2}
	\fmffreeze
	\fmf{photon,left=1}{v2,z1,v3}
	\fmf{photon}{z1,v1}
	\fmf{phantom,tension=2}{h1,z1,v1}
	\fmfdot{v1,z1,v2,v3}
	\end{fmfgraph}}~,\label{eq:SelfEnGr}
\end{align}
which we only include implicitly due to their very similar analytic form as previously computed graphs. In addition we neglect graphs that represent quantum corrections, since they are suppressed in the classical effective action. In particular we neglect the following type of graphs
\bal{	
\raisebox{-0.6cm}{
	\begin{fmfgraph}(50,40)
		\fmfpen{thin}
		\fmfleft{a1,a2}
		\fmfright{b1,b2}
		\fmf{vanilla}{a1,v1,a2}
		\fmf{vanilla}{b1,v2,b2}
		\fmffreeze
		\fmf{photon}{v1,x1,v2}
		\fmfv{decor.shape=circle,decor.filled=shaded,decor.size=0.3w}{x1}
		\fmfdot{v1,v2}
	\end{fmfgraph}}~,}
where the grey blob denotes loop topologies.

\subsection{The double copy of the gauge-theory effective action}
In order to perform the double copy according to \eqref{eq:DCstart} we first need to investigate the independent color structures from the previous section. They are given by 
\begin{align}
\text{LO:} & \; (c\cdot\tilde c), \, (c\cdot c), \, (\tilde c\cdot\tilde c). \nonumber \\
\text{NLO:} & \;  (c\cdot\tilde c)^{2}, \, (c\cdot\tilde c)\, (\tilde c\cdot\tilde c), \,
(\tilde c\cdot c) (c\cdot c), \, (c\cdot c)^{2}, \, (\tilde c\cdot \tilde c)^{2}, 
\\& f^{abc} c_{a} c_{b} c_{c} , \, f^{abc}c_{a}\tilde c_{b}\tilde c_{c}, \,
f^{abc}\tilde c_{a} c_{b} c_{c} , \, f^{abc} \tilde c_{a} \tilde c_{b} \tilde c_{c} ,
\nonumber 
\end{align}
yielding the backbone to write down the terms in \eqref{eq:DCstart}.  We start with the $H$-graph
	\begin{align}
	H:\quad
	\raisebox{-0.6cm}{\begin{fmfgraph*}(50,40)
		\fmfpen{thin}
		\fmfleft{a1,a2}
		\fmfright{b1,b2}
		\fmf{vanilla}{a1,v1,a2}
		\fmf{vanilla}{b1,v2,b2}
		\fmffreeze
		\fmf{photon}{v1,v2}
		\fmfdot{v1,v2}
		\fmfv{label=$\tau_1$}{v1}
		\fmfv{label=$\tilde\tau_2$}{v2}
		\end{fmfgraph*}}\qquad 
	&=\,(2g)^2\left(\frac{i}{\hbar}\right)\,\int d\hat\tau_{1\tilde2}\,\frac{C_H\,N_H}{S_H\,D_H}~,
	\end{align}
	where
	\bal{
          &C_H = (c_{1}\cdot \tilde c_{2}),\quad D_H^{-1}=D_{1\tilde 2},\\
          &S_H=1,\quad N_H =  (p_{1}\cdot \tilde p_{2}).
	}
	Next we have the single-worldline (self-interaction) $D$-graph
	\begin{align}
	D:\quad
	\raisebox{-0.6cm}{
		\begin{fmfgraph*}(50,40)
		\fmfpen{thin}
		\fmfright{a1,a2}
		\fmfleft{b1,b2}
		\fmf{vanilla}{a1,v1,a2}
		\fmf{vanilla}{b1,v2,v5,b2}
		\fmffreeze
		\fmf{photon,right=1}{v2,v5}
		\fmfv{label=$\tau_1$}{v5}
		\fmfv{label=$\tau_2$}{v2}
		\fmfdot{v2,v5}
		\end{fmfgraph*}} \quad
	&= (2g)^2\,\left(\frac{i}{\hbar}\right)\,\int d\hat\tau_{12}\,\frac{C_D\,N_D}{S_D\,D_D}~,
	\end{align}
	where
	\bal{
          &C_D = 	(c_{1}\cdot c_{2}),\quad D_D^{-1}=D_{12},\\
          &S_D = 2,\quad N_D = (p_{1}\cdot p_{2}).
	}
Turning to next-to-leading order contributions, we have the $V$-graph
	\begin{align}
	V:\quad
	&\raisebox{-0.6cm}{
		\begin{fmfgraph*}(50,40)
		\fmfpen{thin}
		\fmfleft{a1,a2}
		\fmfright{b1,b2}
		\fmf{vanilla}{a1,v1,v3,a2}
		\fmf{vanilla}{b1,v2,v4,b2}
		\fmffreeze
		\fmf{photon}{v1,v2}
		\fmf{photon}{v3,v4}
		\fmfdot{v1,v2,v3,v4}
		\fmfv{label=$\tau_1$}{v1}
		\fmfv{label=$\tilde\tau_3$}{v2}
		\fmfv{label=$\tau_2$}{v3}
		\fmfv{label=$\tilde\tau_4$}{v4}
		\end{fmfgraph*}} \quad
	+ \quad
	\raisebox{-0.6cm}{\begin{fmfgraph*}(50,40)
		\fmfpen{thin}
		\fmfleft{a1,a2}
		\fmfright{b1,b2}
		\fmf{vanilla}{a1,v1,a2}
		\fmf{vanilla}{b1,v2,v4,b2}
		\fmffreeze
		\fmf{photon}{v1,v2}
		\fmf{photon}{v1,v4}
		\fmfdot{v1,v2,v4}
		\fmfv{label=$\tau_1$}{v1}
		\fmfv{label=$\tilde\tau_3$}{v2}
		\fmfv{label=$\tilde\tau_4$}{v4}
		\end{fmfgraph*}}\quad
	+\quad
	\raisebox{-0.6cm}{\begin{fmfgraph*}(50,40)
		\fmfpen{thin}
		\fmfright{a1,a2}
		\fmfleft{b1,b2}
		\fmf{vanilla}{a1,v1,a2}
		\fmf{vanilla}{b1,v2,v4,b2}
		\fmffreeze
		\fmf{photon}{v1,v2}
		\fmf{photon}{v1,v4}
		\fmfdot{v1,v2,v4}
		\fmfv{label=$\tilde\tau_3$}{v1}
		\fmfv{label=$\tau_1$}{v2}
		\fmfv{label=$\tau_2$}{v4}
		\end{fmfgraph*}}\quad\nonumber\\[2ex]&
	= (2g)^4\,\left(\frac{i}{\hbar}\right)^2\,\int d\hat\tau_{12\tilde3\tilde4}\,\frac{C_V\,N_V}{S_V\,D_V} ~, 
	\end{align}
	where
	\bal{
		D_V^{-1} &=  D_{1\tilde3}D_{2\tilde4},\quad C_V = (c_{1}\cdot \tilde c_{3})(c_{2}\cdot \tilde c_{4}),\quad S_V = 2,\\
		N_V &= (p_{1}\cdot \tilde p_{3})\,(p_{2}\cdot \tilde p_{4})\, 
		+ \frac{\hbar}{2i}\,\frac{\delta(\tau_{1}-\tau_{2})}{\lambda_2} \, (\tilde p_{3}\cdot \tilde p_{4}) \\
		&\quad + \frac{\hbar}{2i}\, \frac{\delta(\ttau_{3}-\ttau_{4})}{\tilde\lambda_{4}}\, (p_{1}\cdot p_{2})~,
	}
and the $Y$-graph
	\begin{align}
	Y:\quad
	\raisebox{-0.6cm}{
		\begin{fmfgraph*}(50,40)
		\fmfpen{thin}
		\fmfleft{a1,a2}
		\fmfright{b1,b2}
		\fmf{vanilla}{a1,v1,a2}
		\fmf{vanilla}{b1,v2,v3,b2}
		\fmffreeze
		\fmf{photon,tension=1.7}{v1,x1}
		\fmf{photon}{x1,v2}
		\fmf{photon}{x1,v3}
		\fmfdot{v1,v2,v3}
		\fmfv{decor.shape=circle,decor.filled=empty,decor.size=0.12w}{x1}
		\fmfv{label=$\tau_1$}{v1}
		\fmfv{label=$\tilde \tau_2$}{v2}
		\fmfv{label=$\tilde \tau_3$}{v3}
		\fmfv{label=$x$,l.a=90}{x1}
		\end{fmfgraph*}}\quad
	=(2g)^4\,\left(\frac{i}{\hbar}\right)\,\int d\hat\tau_{1\tilde2\tilde3}\,\frac{C_Y\,N_Y}{S_Y\,D_Y}~,
	\end{align}
	where
	\bal{
		&C_Y = f^{abc} c_{1}^{a}\tilde c_{2}^{b}\tilde c_{3}^{c}, \qquad D_Y^{-1}=G_{1\tilde2\tilde3} ,\qquad S_Y=2,\\
		&N_Y=-\frac{1}{2}V^{\mu\nu\rho}_{1\tilde2\tilde3}p_{1\mu}\tilde p_{2\nu}\tilde p_{3\rho}.
	}
\begin{widetext}
Next there is the $C$-graph
	\begin{align}
	C:\quad
	\raisebox{-0.6cm}{
		\begin{fmfgraph*}(50,40)
		\fmfpen{thin}
		\fmfleft{a1,a2}
		\fmfright{b1,b2}
		\fmf{vanilla}{a1,v1,a2}
		\fmf{vanilla}{b1,v2,vn,v4,b2}
		\fmffreeze
		\fmf{photon}{v1,vn}
		\fmf{photon,right=1}{v2,v4}
		\fmfdot{v1,v2,v4,vn}
		\fmfv{label=$\tau_1$}{v1}
		\fmfv{label=$\tilde\tau_4$}{v2}
		\fmfv{label=$\tilde\tau_2$,l.a=135}{vn}
		\fmfv{label=$\tilde\tau_3$}{v4}
                \end{fmfgraph*}}\quad + \quad
	\raisebox{-0.6cm}{
		\begin{fmfgraph*}(50,40)
		\fmfpen{thin}
		\fmfleft{a1,a2}
		\fmfright{b1,b2}
		\fmf{vanilla}{a1,v1,a2}
		\fmf{vanilla}{b1,v2,v4,b2}
		\fmffreeze
		\fmf{photon}{v1,v2}
		\fmf{photon,right=1}{v2,v4}
		\fmfdot{v1,v2,v4}
		\fmfv{label=$\tau_1$}{v1}
		\fmfv{label=$\tilde\tau_2$,l.a=-135}{v2}
		\fmfv{label=$\tilde\tau_3$}{v4}
		\end{fmfgraph*}}\quad
	&= (2g)^4\,\left(\frac{i}{\hbar}\right)^2\,\int d\hat\tau_{1\tilde2\tilde3\tilde4}\,\frac{C_C\,N_C}{S_C\,D_C}~,
	\end{align}
	where
	\bal{
		&C_C= (c_{1}\cdot \tilde c_{2})\,(\tilde c_{3}\cdot \tilde c_{4})\qquad,\qquad D_C^{-1}= D_{1\tilde2}\, D_{\tilde3\tilde4}\qquad,\qquad S_C=2~~,\\
		&N_C = (p_{1}\cdot \tilde p_{2})\,(\tilde p_{3}\cdot \tilde p_{4}) 
		+\frac{\hbar}{i}\,\frac{\delta(\ttau_{2}-\ttau_{4})}{\tilde\lambda_4}\, (p_{1}\cdot \tilde p_{3})~.
	}
Note that here we use the symmetry factor of the first topology of the $C$-graph. This is consistent with our double-copy prescription which states that the correct symmetry factor is the one attributed to the trivalent graph.
        
Next in line is the $B$-graph
	\begin{align}
	B:
	\raisebox{-0.6cm}{
		\begin{fmfgraph*}(50,40)
		\fmfpen{thin}
		\fmfleft{a1,a2}
		\fmfright{b1,b2}
		\fmf{vanilla}{a1,a2}
		\fmf{vanilla}{b1,v2,a,v,k,l,j,v4,r,w,v5,g,h,o,p,u,v6,b2}
		\fmf{photon,left=1,tension=0}{v2,v4}
		\fmf{photon,left=1,tension=0}{v5,v6}
		\fmfdot{v2,v4,v5,v6}
		\fmfv{label=$\ttau_1$}{v4}
		\fmfv{label=$\ttau_4$}{v5}
		\fmfv{label=$\ttau_2$,l.a=15}{v2}
		\fmfv{label=$\ttau_3$,l.a=-15}{v6}
		\end{fmfgraph*}
	} \quad +
	\raisebox{-0.6cm}{
		\begin{fmfgraph*}(50,40)
		\fmfpen{thin}
		\fmfleft{a1,a2}
		\fmfright{b1,b2}
		\fmf{vanilla}{a1,v1,a2}
		\fmf{vanilla}{b1,v2,g,h,j,v4,a,s,d,v5,b2}
		\fmffreeze
		\fmf{photon,left=1,tension=0}{v2,v4}
		\fmf{photon,left=1,tension=0}{v4,v5}
		\fmfdot{v2,v4,v5}
		\fmfv{label=$\ttau_1$}{v4}
		\fmfv{label=$\ttau_2$,l.a=15}{v2}
		\fmfv{label=$\ttau_3$,l.a=-15}{v5}
		\end{fmfgraph*}
	} \quad = (2g)^4\,\left(\frac{i}{\hbar}\right)^2\,\int d\hat\tau_{\tilde1\tilde2\tilde3\tilde4}\,\frac{C_B\,N_B}{S_B\,D_B}~,
	\end{align}
	where
	\bal{
		&C_B = (\tilde c_{1}\cdot \tilde c_{2})\,(\tilde c_{4}\cdot \tilde c_{3})\qquad,\qquad D_B^{-1}= D_{\tilde1\tilde2} D_{\tilde3\tilde4}\qquad,\qquad S_B=8~~,\\
		&N_B = (\tilde p_1 \cdot \tilde p_2) (\tilde p_3 \cdot \tilde p_4)
		+ 2\,\frac{\hbar}{i}\, \frac{\delta(\ttau_{4}-\ttau_{1})}{\tilde\lambda_4}
		(\tilde p_2 \cdot \tilde p_3)~.
	}
Again, the symmetry factor is given by the first topology of the $B$-graph.\\
	The same constructions apply to the mirrored graphs and the graphs mentioned in \eqref{eq:SelfEnGr}. Performing the double copy as discussed in Sec.~\ref{section:dcp} yields our prediction for the exponential of the effective action in dilaton gravity
	\bal{
		\mathcal{\bar M}_{\text{dg}}&= 1 - \kappa^2\,\frac{i}{\hbar}\left(\int d\hat\tau_H\,\frac{N_H\,N_H}{S_H\,D_H} + \int d\hat\tau_D\,\frac{N_D\,N_D}{S_D\,D_D}\right) \\&+ \kappa^4\left(\frac{i}{\hbar}\right)^2\,\left(\int d\hat\tau_V\,\frac{N_V\,N_V}{S_V\,D_V}+\int d\hat\tau_C\,\frac{N_C\,N_C}{S_C\,D_C}+\int d\hat\tau_B\,\frac{N_B\,N_B}{S_B\,D_B}\right) \\
		&\quad+ \kappa^4\, \frac{i}{\hbar}\,\int d\hat\tau_Y\,\frac{N_Y\,N_Y}{S_Y\,D_Y} +\text{(mirrored)}~,\label{eq:copiedaction}
	}
where we introduced the bar notation for $\mathcal{\bar{M}}_\text{dg}$ to point out that it is computed via our double-copy prescription. At this point it is necessary to comment on a subtle feature of the above result. An essential technicality in the previous construction was the notion of pulling apart the gluon-gluon-worldline vertex using a delta function, i.e.~obtaining a trivalent representation. Doing so, our double-copy prescription introduces $\delta(0)$ terms in $\mathcal{\bar{M}}_\text{dg}$. This is a potential hazard but since such terms are of $\mathcal O(\hbar^0)$, they are quantum corrections to $S_{\text{eff,dg}}$ and we neglect them
for the time being.

Another important property of $S_{\text{eff,dg}}$ is that all negative powers of $\hbar$ exponentiate, which is \emph{a priori} not obvious. We check this by taking the logarithm of our double-copy result,  obtaining a perturbative expansion of the effective action,
\begin{align}
\bar{S}_{\text{eff,dg}}& =  S_{\text{pc,free}} + \frac{\hbar}{i}\,\log \mathcal{\bar{M}}_{\text{dg}}= S_{\text{pc,free}} -\kappa^{2}
\int d\hat\tau_{1\tilde2}\,(p_{1}\cdot \tilde p_{2})^{2}\, 
D_{1\tilde2}
-\frac{\kappa^{2}}{2}\int d\hat\tau_{12}
\,(p_{1}\cdot p_{2})^{2}\,  D_{12}
\, \nonumber \\ & +
\frac{\kappa^{4}}{2} \int d\hat\tau_{1\tilde2\tilde3}\,
\left(\frac{1}{2}\,V^{\mu\nu\rho}_{1\tilde2\tilde3}\,p_{1\mu}\,\tilde p_{2\nu}\,\tilde p_{3\rho}\right)^{2}\, G_{1\tilde2\tilde3} \, \nonumber  \\
&+
\frac{\kappa^{4}}{2}\int d\hat\tau_{1\tilde2\tilde3}\,
(p_{1}\cdot \tilde p_{2})\,(p_{1}\cdot \tilde p_{3}) (\tilde p_{3}\cdot \tilde p_{2}) 
D_{1\tilde2}D_{1\tilde3}\, \nonumber  \\ &
+\kappa^{4}\int d\hat\tau_{1\tilde2\tilde3}\,  (p_{1}\cdot \tilde p_{3})
\, (p_{1}\cdot \tilde p_{2}) \,(\tilde p_{3}\cdot \tilde p_{2})
\, D_{1\tilde2}\, D_{\tilde2\tilde3}\nonumber\\&+
\frac{\kappa^4}{2}\int d\hat\tau_{123}\,(p_1\cdot p_2)\,(p_1\cdot p_3)\,(p_2\cdot p_3)\,D_{12}\,D_{13} \nonumber 
\\&+\text{(mirrored)} + \mathcal{O}(\hbar)\, .
\end{align}
\end{widetext}
In the following we will check whether a direct construction within dilaton gravity
reproduces this prediction.

\section{Interaction of point masses in dilaton gravity}
In this section we are performing an analogous analysis as in the previous section for a system of two massive worldlines that are interacting in dilaton gravity. 
\subsection{Dilaton gravity}
The action of dilaton gravity is given by\footnote{Note that the Gibbons-Hawking-York (GHY)
  boundary term \cite{Gibbons:1976ue,York:1972sj} is nonzero also in dimensional regularization
  for asymptotically flat spacetimes, leading to \eqref{dg2}.}
\begin{align}
  \begin{split}
    S_\text{dg} &= - \frac{2}{\kappa^2} \int d^4x \sqrt{-g} \left[ R - 2 \partial_\mu \phi \partial^\mu \phi \right] \nl + (\text{GHY boundary term}) \label{dg1}
  \end{split}\\
  \begin{split}
    &= - \frac{2}{\kappa^2} \int d^4x \sqrt{-g} \big[ g^{\mu\nu} \big( \Gamma^\rho{}_{\mu\lambda} \Gamma^\lambda{}_{\nu\rho} \nl
    - \Gamma^\rho{}_{\mu\nu} \Gamma^\lambda{}_{\rho\lambda} \big)
    - 2 \partial_\mu \phi \partial^\mu \phi \big] , \label{dg2}
  \end{split}
\end{align}
where $\kappa = m_\text{Pl}^{-1} = \sqrt{32\pi G}$ is the gravitational coupling (with Newton constant $G$ and Planck mass $m_\text{Pl}$), $\phi$ is a real scalar field called the dilaton, $R$ is the usual Ricci scalar, $\Gamma^\alpha{}_{\mu\nu} = (\partial_\mu g_{\nu\beta} + \partial_\nu g_{\mu\beta} - \partial_\beta g_{\mu\nu}) g^{\alpha\beta} / 2$ is the Christoffel connection, $g_{\mu\nu}$ is the metric, and $g = \det(g_{\mu\nu})$.  The worldline action of a point mass (pm) coupled to gravity and a dilaton is defined by 
\begin{equation}\label{Spm}
  S_\text{pm} = - m \int d \tau \, e^\phi \sqrt{g_{\mu\nu} u^\mu u^\nu} ,
\end{equation}
with the worldline $x^\mu(\tau)$, the 4-velocity $u^\mu = \dot x^\mu$
and $\phi$, $g_{\mu\nu}$ are evaluated at $x^\mu(\tau)$. We again pass to the first-order version
\begin{align}
  S_\text{pm} &= - \int d\tau \left( p_{\mu} u^\mu - \lambda \left[ e^{-2 \phi} g^{\mu\nu} p_{\mu} p_{\nu} - m^2 \right] \right) .\label{eq:YMWL}
\end{align}
The full action reads
\begin{equation}
S_\text{dg} + S_\text{gf} + S_\text{pm} + \tilde S_\text{pm} ,
\end{equation}
with a gauge fixing part $S_\text{gf}$ specified below.
Note that we do not need an axion field here, which will be important when adding spin
to the point particles \cite{Goldberger:2017ogt}.

\subsection{Weak-field expansion}
The weak-field approximation is defined as the expansion of the full metric around a flat Minkowski background, i.e. 
\begin{align}
  g_{\mu\nu}(x)&= \eta_{\mu\nu} + \kappa h_{\mu\nu}(x) ,\\
  g^{\mu\nu}(x)&=\eta^{\mu\nu} - \kappa h^{\mu\nu}(x) + \kappa^2 h ^{\mu\lambda}(x) h_\lambda^{~\nu}(x) + \mathcal O(\kappa^3) ,\\
\begin{split}
  \sqrt{-g}&=1+\kappa h^\mu_{~\mu}(x) - \frac{\kappa^2}{2} \left(h^{\mu\nu}(x) h_{\mu\nu}(x) - (h^\mu_{~\mu}(x))^2 \right) \nl
  +\mathcal O(\kappa^3) ,
\end{split}
\end{align}
where $\kappa$ is our weak-field (or post-Minkowskian) perturbation parameter and $h_{\mu\nu}$ is the graviton. In principle we are ready to start the computation of the effective action by integrating out the graviton and dilaton field. Nevertheless, at this point one encounters large expressions during intermediate steps, e.g.~the three-graviton vertex will have around 170 terms \cite{DeWitt:1967uc}. However, it is known that one can achieve a high simplification by performing field redefinitions and adding the appropriate gauge fixings \cite{Bern:1999ji}. Here, we aim at a field redefinition that removes the coupling of the dilaton to the worldline and also simplifies the three-graviton vertex to the square of the Yang-Mills one. Our procedure is given by:

First, choosing the following gauge fixing terms to $\mathcal O(\kappa^2)$,
\begin{gather}
  S_\text{gf} = \frac{1}{\kappa^2} \int d^4x \sqrt{-g} f^\mu f_\mu , \\
\begin{split}
  f^\mu &= \Gamma^{\mu}{}_{\nu\sigma} g^{\nu\sigma} + \frac{\kappa^2}{2} \Big[
  -\frac{1}{4} (\partial_\kappa h^{\kappa\lambda}) h_{\lambda}^{~\mu}
  -\frac{1}{4} (\partial^\mu h^{\kappa\lambda}) h_{\kappa\lambda} \nl
  + (\partial^\kappa h^{\mu\lambda}) h_{\kappa\lambda}
  +\frac{3}{16} (\partial^\mu h^\kappa_{~\kappa}) h^\lambda_{~\lambda}
  -\frac{3}{8} (\partial^\kappa h^{\mu}_{~\kappa}) h^\lambda_{~\lambda} \nl
  -\frac{ 3}{8} (\partial^\lambda h^\kappa_{~\kappa})\,h^\mu_{~\lambda}
\Big]~,
\end{split}
\end{gather}
then performing the field redefinitions
\begin{align}
h_{\mu\nu} &\rightarrow h_{\mu\nu} -\eta_{\mu\nu}\left(\frac{1}{2} h^\mu_{~\mu} + 2\phi \right) \nnl
             +\kappa \Big(-\frac{1}{2} h_{\mu\nu} h^\rho_{~\rho}  + \frac{1}{8} \eta_{\mu\nu} h^{\rho}_{~\rho} h^{\sigma}_{~\sigma} + \frac{1}{2} h_{\mu\rho} h^\rho_{~\nu} \nnl
             - 2 \phi h_{\mu\nu} + 2 \phi^2 \eta_{\mu\nu}+\phi h_{\mu\nu} h^{\rho}_{~\rho} \Big) , \\
\phi &\rightarrow \phi + \frac{1}{4}\,h^\mu_{~\mu}~,
\end{align}
and finally, adding the total derivative
\begin{equation}
\begin{split}
0 &= S_\text{TD} = \int d^4 x \Big[ \partial_\mu( (\partial_\nu h^{\mu\kappa}) h^\nu_{~\kappa}) - \partial_\mu (h^{\mu\nu} (\partial_\kappa h^\kappa_{~\nu}))
\\
&\quad + \kappa \big(\frac{1}{4} \partial_\mu(h^{\mu\nu}(\partial_\nu h^{\sigma\nu}) h_{\sigma\nu}) -\frac{1}{4} \partial_\mu ((\partial_\nu h^{\mu\lambda})h_{\lambda\kappa} h^{\kappa\nu})
\\
&\quad -\frac{1}{4} \partial_\mu (h^{\mu\nu} (\partial_\lambda h_{\nu\rho})h^{\rho\lambda})+ \frac{1}{4} \partial_\mu (h^\mu_{~\nu} h^{\nu\lambda} (\partial_\sigma h_\lambda^{~\sigma}))\big)\Big] ,
\end{split}
\end{equation}
to the action. 
This procedure decouples the point masses on the worldline from the dilaton,
\begin{align}
  S_\text{pm} &= - \int d\tau \bigg( p_{\mu} u^\mu - \lambda \bigg[ \bigg( \eta^{\mu\nu}  - \kappa h^{\mu\nu} + \frac{\kappa^2}{2} h^{\mu}{}_{\rho} h^{\rho\nu} \bigg) p_{\mu} p_{\nu}\nnl
                - m^2 \bigg] \bigg) + \Order(\kappa^3) \label{eq:GRWL},
\end{align}
which means we only need to worry about the graviton-dependent part of the field action for the computation of the classical effective action $S_{\text{eff,dg}}$ at next-to-leading order. Furthermore we observe a considerable simplification in the three-graviton interaction, i.e.~it is given by 
\begin{align}
  &S_\text{dg} + S_\text{gf} + S_{TD} \\
  &= \int d^4x \bigg[ \frac{1}{2} \partial_\rho h_{\mu\nu} \partial^\rho h^{\mu\nu}
                    + \frac{\kappa}{4}\Big(h_{\mu\nu} \partial^\mu \partial^\nu h_{\rho\sigma} h^{\rho\sigma} \nnl
                    + 2 h_{\mu\nu} \partial^\sigma h^{\mu}_{~\rho} \partial^\nu h^{\rho}_{~\sigma}
  - h_{\mu\nu} \partial^\sigma h_\rho{}^\mu \partial^\rho h^\nu_{~\sigma}
    - h_{\rho\sigma} \partial^\rho h_{\mu\nu} \partial^\sigma h^{\mu\nu} \nnl
    - \partial_\rho \partial_\sigma h_{\mu\nu} h^{\rho\mu} h^{\sigma\nu} \Big)
    \bigg] + \Order(\kappa^2, \phi) \\
  &=  \int d^4x \bigg[ \frac{1}{2} \partial_\rho h_{\mu\nu} \partial^\rho h^{\mu\nu}
    + \frac{\kappa}{4 \cdot 3!} V_{123}^{\mu\alpha\gamma} V_{123}^{\nu\beta\delta} h_{1\mu\nu} h_{2\alpha\beta} h_{3\gamma\delta} \bigg] \nnl
    + \Order(\kappa^2, \phi) , \label{Sdgexpand}
\end{align}
where $V_{123}^{\mu\alpha\gamma}$ is again the color independent part of the three gluon interaction defined by \eqref{defV}. We also introduced fiducial indices 1, 2, 3, on $h_{\mu\nu}$ to
indicate on which $h_{\mu\nu}$ the partial derivatives in $V_{123}^{\mu_1\mu_2\mu_3}$ are applied.
The propagator therefore reads
\bal{
	\langle h_{\mu\nu}(x)\, h_{\rho\sigma}(y)\rangle_{0}\, =\,  \frac{\hbar}{i}\, \eta_{\mu(\rho}\eta_{\sigma)\nu}\, D(x-y)~,
}
where the round brackets on the indices indicate a symmetrization of unit weight,
\bal{
\eta_{\mu(\rho}\eta_{\sigma)\nu} = \frac{1}{2}\left(\eta_{\mu\rho}\,\eta_{\sigma\nu}+\eta_{\mu\sigma}\,\eta_{\rho\nu}\right)~.
}

\subsection{Effective action of gravitating particles to next-to-leading order}
Similar to the Yang-Mills case in Sec.~\ref{YMTECH}, we compute $S_{\text{eff,dg}}$ by integrating out the graviton and scalar fields,
\begin{multline}
  e^{\frac{i}{\hbar} S_\text{eff,dg}} =  e^{\frac{i}{\hbar} S_\text{pm,free}} \mathcal{M}_\text{dg} \\
  = \text{const} \int \mathcal{D} h \mathcal{D} \phi \, e^{\frac{i}{\hbar} ( S_\text{dg} + S_\text{gf} + S_\text{pm} + \tilde S_\text{pm} )} .
\end{multline}
Again the (divergent) normalization constant on the right-hand side is chosen
such that $\mathcal{M}_\text{dg} = 1$ for $\kappa \rightarrow 0$ and
$S_\text{pm,free}$ is given by $S_\text{pm}+\tilde{S}_\text{pm}$ for $\kappa \rightarrow 0$.
We also neglect quantum corrections and keep all mirrored graphs implicit.

The relevant classical diagrams for $\mathcal{M}_\text{dg}$ 
to order $\kappa^4$ read
\begin{align}  
\raisebox{-0.6cm}{
\begin{fmfgraph*}(50,40)
	\fmfpen{thin}
	\fmfleft{a1,a2}
	\fmfright{b1,b2}
	\fmf{vanilla}{a1,v1,a2}
	\fmf{vanilla}{b1,v2,b2}
	\fmffreeze
	\fmf{dbl_wiggly}{v1,v2}
	\fmfdot{v1,v2}
	\fmfv{label=$\tau_1$}{v1}
	\fmfv{label=$\ttau_2$}{v2}
      \end{fmfgraph*}} \quad &= - \frac{i \kappa^2}{\hbar} \int d\hat\tau_{1\tilde2} \, (p_1 \cdot \tilde p_2)^2 D_{1\tilde2} ~, \\[2ex]
  \raisebox{-0.6cm}{
	\begin{fmfgraph*}(50,40)
	\fmfpen{thin}
	\fmfright{a1,a2}
	\fmfleft{b1,b2}
	\fmf{vanilla}{a1,v1,a2}
	\fmf{vanilla}{b1,v2,v5,b2}
	\fmffreeze
	\fmf{dbl_wiggly,right=1}{v2,v5}
	\fmfv{label=$\tau_1$}{v5}
	\fmfv{label=$\tau_2$}{v2}
	\fmfdot{v2,v5}
	\end{fmfgraph*}} \quad &= -\frac{i\,\kappa^2}{2\,\hbar}\,\int d\hat\tau_{12}\,(p_1\cdot p_2)^2\,D_{12}~,
\end{align}
\begin{widetext}
\begin{align}
\raisebox{-0.6cm}{
\begin{fmfgraph*}(50,40)
	\fmfpen{thin}
	\fmfleft{a1,a2}
	\fmfright{b1,b2}
	\fmf{vanilla}{a1,v1,v3,a2}
	\fmf{vanilla}{b1,v2,v4,b2}
	\fmffreeze
	\fmf{dbl_wiggly}{v1,v2}
	\fmf{dbl_wiggly}{v3,v4}
	\fmfdot{v1,v2,v3,v4}
	\fmfv{label=$\tau_1$}{v1}
	\fmfv{label=$\ttau_2$}{v2}
	\fmfv{label=$\ttau_3$}{v4}
	\fmfv{label=$\tau_4$}{v3}
      \end{fmfgraph*}} \quad &= \frac{i^2 \kappa^4}{2\hbar^2} \int d\hat\tau_{1\tilde2\tilde3 4} \, (p_1 \cdot \tilde p_2)^2 (p_4 \cdot \tilde p_3)^2 D_{1\tilde2} D_{4\tilde3} ~, \\[2ex]
\raisebox{-0.6cm}{
\begin{fmfgraph*}(50,40)
	\fmfpen{thin}
	\fmfleft{a1,a2}
	\fmfright{b1,b2}
	\fmf{vanilla}{a1,v1,a2}
	\fmf{vanilla}{b1,v2,v4,v5,v6,b2}
	\fmffreeze
	\fmf{dbl_wiggly}{v1,v2}
	\fmf{dbl_wiggly}{v1,v6}
	\fmfdot{v1,v2,v6}
	\fmfv{label=$\tau_1$}{v1}
	\fmfv{label=$\ttau_2$}{v2}
	\fmfv{label=$\ttau_3$}{v6}
\end{fmfgraph*}} \quad &= \frac{i \kappa^4}{2 \hbar} \int d\hat\tau_{1\tilde2\tilde3} \, (p_1 \cdot \tilde p_2) (p_1 \cdot \tilde p_3) (\tilde p_2 \cdot \tilde p_3) D_{1\tilde2} D_{1\tilde3}~, \\[2ex]
\raisebox{-0.6cm}{
\begin{fmfgraph*}(50,40)
	\fmfpen{thin}
	\fmfleft{a1,a2}
	\fmfright{b1,b2}
	\fmf{vanilla}{a1,v1,a2}
	\fmf{vanilla}{b1,v2,v3,v5,b2}
	\fmffreeze
	\fmf{dbl_wiggly,tension=1.7}{v1,x1}
	\fmf{dbl_wiggly}{x1,v2}
	\fmf{dbl_wiggly}{x1,v5}
	\fmfdot{v1,v2,v5}
	\fmfv{decor.shape=circle,decor.filled=empty,decor.size=0.12w}{x1}
	\fmfv{label=$x$,l.a=90}{x1}
	\fmfv{label=$\tau_1$}{v1}
	\fmfv{label=$\ttau_2$}{v2}
	\fmfv{label=$\ttau_3$}{v5}
      \end{fmfgraph*}} \quad &= \frac{i \kappa^4}{8 \hbar} \int d\hat\tau_{1\tilde2\tilde3} \, \left ( V_{1\tilde2\tilde3}^{\mu\nu\rho} p_{1\mu} \tilde p_{2\nu} \tilde p_{3\rho} \right )^2 G_{1\tilde2\tilde3}~,\\[2ex]
  \raisebox{-0.6cm}{
  	\begin{fmfgraph*}(50,40)
  	\fmfpen{thin}
  	\fmfleft{a1,a2}
  	\fmfright{b1,b2}
  	\fmf{vanilla}{b1,b2}
  	\fmf{vanilla}{a1,v2,a,v,k,l,j,v4,r,w,v5,g,h,o,p,u,v6,a2}
  	\fmf{dbl_wiggly,right=1,tension=0}{v2,v4}
  	\fmf{dbl_wiggly,right=1,tension=0}{v5,v6}
  	\fmfdot{v2,v4,v5,v6}
  	\fmfv{label=$\tau_1$}{v4}
  	\fmfv{label=$\tau_4$}{v5}
  	\fmfv{label=$\tau_2$,l.a=180}{v2}
  	\fmfv{label=$\tau_3$,l.a=180}{v6}
  	\end{fmfgraph*}
  } ~~ &= \frac{i^2 \kappa^4}{2^3 \hbar^2} \int d\hat\tau_{1234} \, (p_1 \cdot p_2)^2 (p_4 \cdot p_3)^2 D_{12} D_{43}~ , \\[4ex]
  \raisebox{-0.6cm}{
  	\begin{fmfgraph*}(50,40)
  	\fmfpen{thin}
  	\fmfleft{a1,a2}
  	\fmfright{b1,b2}
  	\fmf{vanilla}{b1,v1,b2}
  	\fmf{vanilla}{a1,v2,g,h,j,v4,a,s,d,v5,a2}
  	\fmffreeze
  	\fmf{dbl_wiggly,right=1,tension=0}{v2,v4}
  	\fmf{dbl_wiggly,right=1,tension=0}{v4,v5}
  	\fmfdot{v2,v4,v5}
  	\fmfv{label=$\tau_1$}{v4}
  	\fmfv{label=$\tau_2$,l.a=180}{v2}
  	\fmfv{label=$\tau_3$,l.a=180}{v5}
  	\end{fmfgraph*}
  } ~~ &= \frac{i \kappa^4}{2 \hbar} \int d\hat\tau_{123} \, (p_1 \cdot p_2) (p_1 \cdot p_3) (p_2 \cdot p_3) D_{12} D_{13} ~ ,\\[2ex]
	\raisebox{-0.6cm}{
  \begin{fmfgraph*}(50,40)
  \fmfpen{thin}
  \fmfleft{a1,a2}
  \fmfright{b1,b2}
  \fmf{vanilla}{a1,v1,a,s,d,a2}
  \fmf{vanilla}{b1,vn,v2,v3,v4,b2}
  \fmffreeze
  \fmf{dbl_wiggly}{v1,vn}
  \fmf{dbl_wiggly,left=1}{v2,v4}
  \fmfdot{v1,v2,v4,vn}
  \fmfv{label=$\tau_2$}{v1}
  \fmfv{label=$\ttau_3$}{v4}
  \fmfv{label=$\ttau_1$}{vn}
  \fmfv{label=$\ttau_4$}{v2}
  \end{fmfgraph*}
} \quad &= \frac{i^2 \kappa^4}{2 \hbar^2} \int d\hat\tau_{\tilde12\tilde3\tilde4} \, (\tilde p_1 \cdot p_2)^2 (\tilde p_4 \cdot \tilde p_3)^2 D_{\tilde12} D_{\tilde4\tilde3} , \\[2ex]
\raisebox{-0.6cm}{
\begin{fmfgraph*}(50,40)
\fmfpen{thin}
\fmfleft{a1,a2}
\fmfright{b1,b2}
\fmf{vanilla}{a1,v1,a,s,a2}
\fmf{vanilla}{b1,v2,v4,v5,b2}
\fmffreeze
\fmf{dbl_wiggly}{v1,v2}
\fmf{dbl_wiggly,left=1}{v2,v5}
\fmfdot{v1,v2,v5}
\fmfv{label=$\tau_2$}{v1}
\fmfv{label=$\ttau_3$}{v5}
\fmfv{label=$\ttau_1$}{v2}
\end{fmfgraph*}
} \quad &= \frac{i \kappa^4}{\hbar} \int d\hat\tau_{\tilde12\tilde3} \, (\tilde p_1 \cdot p_2) (\tilde p_1 \cdot \tilde p_3) (p_2 \cdot \tilde p_3) D_{\tilde12} D_{\tilde1\tilde3}~ .
\end{align}
Summarizing, we find upon taking the logarithm that the effective action in dilaton gravity in a graphical notation reads
\begin{align}
S_\text{eff,dg}  =&  S_{\text{pm,free}} + \frac{\hbar}{i}\log \mathcal{M}_{dg} =  S_{\text{pm,free}} +
\raisebox{-0.6cm}{
		\begin{fmfgraph}(50,40)
		\fmfpen{thin}
		\fmfleft{a1,a2}
		\fmfright{b1,b2}
		\fmf{vanilla}{a1,v1,a2}
		\fmf{vanilla}{b1,v2,b2}
		\fmffreeze
		\fmf{dbl_wiggly}{v1,v2}
		\fmfdot{v1,v2}
		\fmfv{label=$\tau_1$}{v1}
		\fmfv{label=$\ttau_2$}{v2}
		\end{fmfgraph}} \quad +\quad 
		\raisebox{-0.6cm}{
		\begin{fmfgraph}(50,40)
		\fmfpen{thin}
		\fmfleft{a1,a2}
		\fmfright{b1,b2}
		\fmf{vanilla}{a1,v1,a2}
		\fmf{vanilla}{b1,v2,v4,v5,v6,b2}
		\fmffreeze
		\fmf{dbl_wiggly}{v1,v2}
		\fmf{dbl_wiggly}{v1,v6}
		\fmfdot{v1,v2,v6}
			\end{fmfgraph}} 
				\quad +\quad
					\raisebox{-0.6cm}{
		\begin{fmfgraph}(50,40)
		\fmfpen{thin}
		\fmfleft{a1,a2}
		\fmfright{b1,b2}
		\fmf{vanilla}{a1,v1,a2}
		\fmf{vanilla}{b1,v2,v3,v5,b2}
		\fmffreeze
		\fmf{dbl_wiggly,tension=1.7}{v1,x1}
		\fmf{dbl_wiggly}{x1,v2}
		\fmf{dbl_wiggly}{x1,v5}
		\fmfdot{v1,v2,v5}
		\fmfv{decor.shape=circle,decor.filled=empty,decor.size=0.12w}{x1}
		\fmfv{label=$x$,l.a=90}{x1}
		\fmfv{label=$\tau_1$}{v1}
		\fmfv{label=$\ttau_2$}{v2}
		\fmfv{label=$\ttau_3$}{v5}
		\end{fmfgraph}} 
		\quad \nonumber\\[2ex]
	&+\quad 
		\raisebox{-0.6cm}{
		\begin{fmfgraph}(50,40)
		\fmfpen{thin}
		\fmfleft{a1,a2}
		\fmfright{b1,b2}
		\fmf{vanilla}{a1,v1,a,s,a2}
		\fmf{vanilla}{b1,v2,v4,v5,b2}
		\fmffreeze
		\fmf{dbl_wiggly}{v1,v2}
		\fmf{dbl_wiggly,left=1}{v2,v5}
		\fmfdot{v1,v2,v5}
		\fmfv{label=$\tau_2$}{v1}
		\fmfv{label=$\ttau_3$}{v5}
		\fmfv{label=$\ttau_1$}{v2}
		\end{fmfgraph}
	} \quad +\quad
	\raisebox{-0.6cm}{
		\begin{fmfgraph}(50,40)
		\fmfpen{thin}
		\fmfleft{a1,a2}
		\fmfright{b1,b2}
		\fmf{vanilla}{b1,v1,b2}
		\fmf{vanilla}{a1,v2,g,h,j,v4,a,s,d,v5,a2}
		\fmffreeze
		\fmf{dbl_wiggly,right=1,tension=0}{v2,v4}
		\fmf{dbl_wiggly,right=1,tension=0}{v4,v5}
		\fmfdot{v2,v4,v5}
		\fmfv{label=$\tau_1$}{v4}
		\fmfv{label=$\tau_2$,l.a=180}{v2}
		\fmfv{label=$\tau_3$,l.a=180}{v5}
		\end{fmfgraph}}
	+\quad
	\raisebox{-0.6cm}{
		\begin{fmfgraph}(50,40)
		\fmfpen{thin}
		\fmfright{a1,a2}
		\fmfleft{b1,b2}
		\fmf{vanilla}{a1,v1,a2}
		\fmf{vanilla}{b1,v2,v5,b2}
		\fmffreeze
		\fmf{dbl_wiggly,right=1}{v2,v5}
		\fmfv{label=$\tau_1$}{v5}
		\fmfv{label=$\tau_2$}{v2}
		\fmfdot{v2,v5}
		\end{fmfgraph}} + (\text{mirrored})\, . \label{Seffdg}
\end{align}
\end{widetext}
Now that we have obtained all classical contributions from integrating out the graviton and dilaton, we can compare this result with that from the double copy \eqref{eq:copiedaction}. Going through the expression term by term we indeed find 
\bal{
\bar S_{\text{eff,dg}} = S_{\text{eff,dg}} + \int d\tau \, i( \psi^\dagger \dot \psi + \tilde \psi^\dagger \dot{\tilde \psi}) ~,
}
where the last term comes from the difference between $S_{\text{pc,free}}$ and $S_{\text{pm,free}}$.
However, the dynamics of the auxiliary field $\psi$ is now decoupled and trivial such that it may be dropped.
Therefore we conclude that our double-copy prescription \eqref{eq:DCprescription} yields the correct classical effective action up to next-to-leading order in the weak-field expansion.

\section{Post-Newtonian evaluation of the effective action}
In this section, we evaluate the effective action in the post-Newtonian approximation.
After adopting a gauge for the worldline parameter, the equations of motion for the
Lagrange multipliers $\lambda$ and the energies $p^0$ of the particles become
algebraic and these variables can be eliminated from the action. This leads to
an action in Hamiltonian form, which agrees with previous results in scalar-tensor theory
\cite{Damour:1992we}, that includes dilaton gravity as a special case.

The post-Newtonian approximation is a refinement of the weak-field approximation to bound
binaries. The third Kepler law, or the virial theorem, then tells us that
(with equality in the circular orbit case)
\begin{equation}
  \frac{\vct{v}_r^2}{c^2} \sim \frac{\kappa^2 (m+\tilde{m})}{c^2 32\pi r} ,
\end{equation}
where $r = | \vct{x} - \tilde{\vct{x}} |$, $\vct{v}_r$ is the relative velocity of the two particles,
and we have restored the speed of light $c$. The weak-field approximation
therefore implies small velocities in the case of bound binaries. It is convenient to use $c^{-1}$ as
a formal post-Newtonian counting parameter, such that $\kappa = \Order(c^{-1})$ as well as
\begin{gather}
  (u^\mu) = (1, \vct{v}) = (\Order(c^0), \Order(c^{-1})) , \\
  (p_\mu) = (E, - \vct{p}) = (\Order(c^0), \Order(c^{-1})) ,
\end{gather}
$\lambda = \Order(c^0)$ and $\partial_t = \Order(c^{-1})$. The post-Newtonian expansion of the propagator
then reads
\begin{align}\label{propPN}
D(x) &= \int \frac{d^4 k}{(2\pi)^4} \frac{e^{-i k_\mu x^\mu}}{k_\mu k^\mu + i \epsilon} \nonumber \\
&= - \int \frac{d^3 \vct{k}}{(2\pi)^3} \frac{e^{i \vct{k} \cdot \vct{x}}}{\vct{k}^2} \left[ 1 - \frac{\partial_t^2}{\vct{k}^2} + \frac{\partial_t^4}{\vct{k}^4} + \dots \right] \delta(t) .
\end{align}

Recall that the effective action is invariant under reparametrizations  of the worldline
parameters. We may therefore pick a gauge for them. In the post-Newtonian approximation, it
is useful to fix $\tau = t = \text{coordinate time}$.
Then the effective action is the sum of the free terms $S_{\text{pm,free}}$
and the connected graphs shown in \eqref{Seffdg}. Most of these graphs are actually
vanishing self-interactions, see Appendix \ref{self}. This leaves us with
(stripping off the overall factors $i/\hbar$)
\begin{align}
\begin{split}
\raisebox{-0.4cm}{
\begin{fmfgraph*}(50,30)
	\fmfpen{thin}
	\fmfleft{a1,a2}
	\fmfright{b1,b2}
	\fmf{vanilla}{a1,v1,a2}
	\fmf{vanilla}{b1,v2,b2}
	\fmffreeze
	\fmf{dbl_wiggly}{v1,v2}
	\fmfdot{v1,v2}
      \end{fmfgraph*}} &= \frac{\kappa^2}{4\pi} \int dt \, \frac{\lambda \tilde \lambda E^2 \tilde E^2}{2 r}
                         \bigg[ 2 - 4 \frac{\vct{p} \cdot \tilde{\vct{p}}}{E \tilde E} \nl
                         + \vct{v} \cdot \tilde{\vct{v}}
                         - (\vct{n} \cdot \vct{v}) (\vct{n} \cdot \tilde{\vct{v}}) \bigg]
                       + \Order(c^{-6}) , \label{onegravPN}
\end{split}\\[2ex]
\raisebox{-0.4cm}{
\begin{fmfgraph*}(50,40)
	\fmfpen{thin}
	\fmfleft{a1,a2}
	\fmfright{b1,b2}
	\fmf{vanilla}{a1,v1,a2}
	\fmf{vanilla}{b1,v2,v4,v5,v6,b2}
	\fmffreeze
	\fmf{dbl_wiggly}{v1,v2}
	\fmf{dbl_wiggly}{v1,v6}
	\fmfdot{v1,v2,v6}
\end{fmfgraph*}} &= \frac{\kappa^4}{(4\pi)^2} \int dt \frac{\lambda \tilde \lambda^2 E^2 \tilde E^4}{2 r^2} + \Order(c^{-6}) , \\[2ex]
\raisebox{-0.4cm}{
\begin{fmfgraph*}(50,40)
	\fmfpen{thin}
	\fmfleft{a1,a2}
	\fmfright{b1,b2}
	\fmf{vanilla}{a1,v1,a2}
	\fmf{vanilla}{b1,v2,v3,v5,b2}
	\fmffreeze
	\fmf{dbl_wiggly,tension=1.7}{v1,x1}
	\fmf{dbl_wiggly}{x1,v2}
	\fmf{dbl_wiggly}{x1,v5}
	\fmfdot{v1,v2,v5}
	\fmfv{decor.shape=circle,decor.filled=empty,decor.size=0.12w}{x1}
\end{fmfgraph*}} &= 0 + \Order(c^{-6}) ,
\end{align}
where $\vct{n} = (\vct{x} - \tilde{\vct{x}}) / r$, and we used
\begin{equation}
  \int \frac{d^d \vct{k}}{(2\pi)^d} \frac{e^{i \vct{k} \cdot \vct{x}}}{(\vct{k}^2)^\alpha}
  = \frac{1}{(4\pi)^{d/2}} \frac{\Gamma(d/2 - \alpha)}{\Gamma(\alpha)} \left( \frac{\vct{x}^2}{4} \right)^{\alpha - d/2} . 
\end{equation}
Moreover in \eqref{onegravPN} we anticipated that $\lambda = \text{const} + \Order(c^{-2}) = E$,
allowing us to drop time derivatives of $\lambda$, $E$. The effective action in the post-Newtonian approximation
finally reads
\begin{equation}
  \begin{split}
  S_\text{eff,dg} &= \int dt \bigg[ \vct{p} \cdot \vct{v} - E + \lambda ( E^2 - m^2 - \vct{p}^2 ) \nl
    + \tilde{\vct{p}} \cdot \tilde{\vct{v}} - \tilde E + \tilde \lambda ( \tilde E^2 - \tilde m^2 - \tilde{\vct{p}}^2 ) \nl
  + \frac{\kappa^2 \lambda \tilde \lambda E^2 \tilde E^2}{8\pi r}
                         \left( 2 - 4 \frac{\vct{p} \cdot \tilde{\vct{p}}}{E \tilde E}
                         + \vct{v} \cdot \tilde{\vct{v}}
                         - (\vct{n} \cdot \vct{v}) (\vct{n} \cdot \tilde{\vct{v}}) \right) \nl
  + \frac{\kappa^4 \lambda \tilde \lambda^2 E^2 \tilde E^4}{2 (4\pi r)^2} + \frac{\kappa^4 \tilde \lambda \lambda^2 \tilde E^2 E^4}{2 (4\pi r)^2}
  + \Order(c^{-6})
\bigg] .
\end{split}
\end{equation}

Next, we vary the effective action with respect to $\lambda$ and $E$ to arrive at
\begin{align}
\begin{split}
  0 &= E^2 - m^2 - \vct{p}^2 \nl
  + \frac{\kappa^2 \tilde \lambda E^2 \tilde E^2}{8\pi r}
      \left[ 2 - 4 \frac{\vct{p} \cdot \tilde{\vct{p}}}{E \tilde E} + \vct{v} \cdot \tilde{\vct{v}} - (\vct{n} \cdot \vct{v}) (\vct{n} \cdot \tilde{\vct{v}}) \right] \nl
      + \frac{\kappa^4}{(4\pi)^2} \frac{\tilde \lambda^2 E^2 \tilde E^4 + 2 \lambda \tilde \lambda E^4 \tilde E^2}{2r^2}
      + \Order(c^{-6}) ,
\end{split}\\
  0 &= - 1 + 2 \lambda E + \frac{\kappa^2 \lambda \tilde \lambda E \tilde E^2}{2\pi r} + \Order(c^{-4}) ,
\end{align}
and similar for the tilded variables. Solving iteratively for $\lambda, E$, using also
$\vct{p} = m \vct{v} + \Order(c^{-3})$,
\begin{align}
  E &= m + \frac{\vct{p}^2}{2 m} - \frac{\kappa^2 m \tilde m}{16\pi r} - \frac{\vct{p}^4}{8 m^3}
        + \frac{2 \kappa^4 m \tilde m (2 m + \tilde m)}{(32\pi r)^2} \nl
        - \frac{\kappa^2 m \tilde m}{32\pi r} \left[ \frac{\vct{p}^2}{m^2}
        + \frac{\tilde{\vct{p}}^2}{\tilde m^2} - \frac{3 (\vct{p} \cdot \tilde{\vct{p}})}{m \tilde m}
        - \frac{(\vct{n} \cdot \vct{p}) (\vct{n} \cdot \tilde{\vct{p}})}{m \tilde m} \right] \nnl
      + \Order(c^{-6}), \\
  \lambda &= \frac{1}{2 m} \left[ 1 - \frac{\vct{p}^2}{2 m^2} - \frac{\kappa^2 \tilde m}{16\pi r} \right] + \Order(c^{-4}) ,
\end{align}
and substituting these solutions into the action, we arrive at (recall $\kappa^2 = 32\pi G$)
\begin{align}
  S_\text{eff,dg} &= \int dt \left[ \vct{p} \cdot \vct{v} + \tilde{\vct{p}} \cdot \tilde{\vct{v}} - H \right] , \\
    H &= m + \tilde m + \frac{\vct{p}^2}{2 m} + \frac{\tilde{\vct{p}}^2}{2 \tilde m}
    - \frac{2G m \tilde m}{r} \nnl
  - \frac{\vct{p}^4}{8 m^3} - \frac{\tilde{\vct{p}}^4}{8 \tilde m^3}
  + \frac{2 G^2 m \tilde m (m + \tilde m)}{r^2} \nnl
  - \frac{G m \tilde m}{r} \left[ \frac{\vct{p}^2}{m^2} + \frac{\tilde{\vct{p}}^2}{\tilde m^2}
    - \frac{3 \vct{p} \cdot \tilde{\vct{p}}}{m \tilde m}
    - \frac{(\vct{n} \cdot \vct{p}) (\vct{n} \cdot \tilde{\vct{p}})}{m \tilde m}
  \right] \nnl
  + \Order(c^{-6}) .
\end{align}
Notice the extra factor of $2$ in the Newtonian potential, which is due
to the dilaton. Throughout this derivation, higher-order time derivatives (like
accelerations) are removed by inserting the respective equations of motion. This
corresponds to a variable redefinition at the considered order \cite{Damour:1990jh}.
[In particular, this justifies the dropping of time derivatives of $\lambda$, $E$
and the use of $\vct{p} = m \vct{v} + \Order(c^{-3})$ above.] Further note that $H$ is the Hamiltonian
describing the motion of the binary system in dilaton gravity.

We may compare our result to scalar-tensor theory \cite{Damour:1992we}.
Scalar-tensor theories (in the so-called Einstein frame) are based on
the same field action as dilaton gravity \eqref{dg1}. However, the source
part \eqref{Spm} is more generic in scalar-tensor theories,
\begin{equation}
  \begin{split}
  S_\text{pm,st} &= - m \int d \tau \, \sqrt{g_{\mu\nu} u^\mu u^\nu} \nl
  \times \left[ 1 + \alpha_0 \phi + \frac{1}{2} (\alpha_0^2+\beta_0) \phi^2 + \Order(\kappa^3) \right] ,
  \end{split}
\end{equation}
with the so-called sensitivities $\alpha_0$ and $\beta_0$.
Comparing to \eqref{Spm} we find that $\alpha_0 = 1$ and $\beta_0 = 0$ for dilaton gravity.
The Hamiltonian for scalar-tensor gravity, following via a Legendre
transform from the Lagrangian stated in Eq.~(3.7) of Ref.~\cite{Damour:1992we}, reads
\begin{align}
  H &= m + \tilde m + \frac{\vct{p}^2}{2 m} +  \frac{\tilde{\vct{p}}^2}{2 \tilde m} - \frac{\tilde G m \tilde m}{r}
  - \frac{\vct{p}^4}{8 m^3} - \frac{\tilde{\vct{p}}^4}{8 \tilde m^3} \nnl
  - \frac{\tilde G m \tilde m}{r} \left[ \frac{3 \vct{p}^2}{2 m^2} + \frac{3 \tilde{\vct{p}}^2}{2 \tilde m^2}
    - \frac{7 \vct{p} \cdot \tilde{\vct{p}}}{2 m \tilde{m}}
    - \frac{(\vct{n} \cdot \vct{p}) (\vct{n} \cdot \tilde{\vct{p}})}{2 m \tilde m} \right] \nnl
  + \frac{2 G m \alpha_0 \tilde m \tilde \alpha_0}{r} \left[ \frac{\vct{p}}{m} - \frac{\tilde{\vct{p}}}{\tilde m} \right]^2
  + \frac{\tilde G^2 m \tilde m}{2 r^2} (m + \tilde m) \nnl
  + \frac{G^2 m \tilde m}{2 r^2} ( m \alpha_0^2 \tilde \beta_0 + \tilde m \tilde \alpha_0^2 \beta_0 )
  + \Order(c^{-6}) ,
\end{align}
where $\tilde G = G (1 + \alpha_0 \tilde \alpha_0)$.
This Hamiltonian agrees with ours for $\alpha_0 = 1$ and $\beta_0 = 0$, as expected.
Note that for $\alpha_0 = 0 = \beta_0$ one obtains the result of general relativity.

In closing let us compare the complexities of establishing the result above with the
more traditional approach of Ref.~\cite{Damour:1992we}. At this low order in perturbation
theory it is hard to claim a clear superiority of the double-copy approach.
We regard our work as a proof of concept, which we expect to simplify calculations at
higher orders dramatically, as it is the case in scattering amplitudes.
But already at the next-to-leading order established here, our approach allows one to obtain the
post-Minkowskian Feynman integrals in a simple way, where in a more traditional
approach the three-graviton vertex contains around 170 terms \cite{DeWitt:1967uc}.

\section{Conclusions}
In the present paper, we proposed an adaption of the BCJ double copy
\cite{Bern:2008qj, Bern:2010yg, Bern:2010ue} to the classical effective
action of binary systems and demonstrated its validity to
next-to-leading order in the weak-field/post-Minkowskian approximation. This is another application
for a BCJ-like double copy beyond (quantum) scattering amplitudes, next to
classical solutions for the field equations \cite{Monteiro:2014cda,Luna:2015paa,Luna:2016hge}
and weak-field approximations for the radiation from classical binary point-sources
\cite{Shen:2018ebu,Goldberger:2016iau,Goldberger:2017vcg,Goldberger:2017ogt,Goldberger:2017frp,Chester:2017vcz}. The former application of the double copy operates at the level
of the gauge field and metric which are gauge-covariant quantities, unlike
gauge-invariant scattering amplitudes.
Similarly, in the present work, we formulated a double copy between effective actions, which are not
\emph{manifestly} gauge independent either: the actions depend on gauge-dependent canonical
momenta, and on the dilaton-gravity side the positions are gauge dependent,
too.\footnote{The (conservative) effective actions encode the binding energy (energy levels in
  a quantum mechanical analog), which are gauge independent. However, the variables (like momenta)
  that are used when the double copy is performed are not gauge invariant.}
This provides growing evidence that the BCJ double copy can be adapted to gauge-covariant quantities,
which, if true, would liberate it to a considerably larger realm of applications.
Furthermore, our work is an adaptation of the BCJ double copy to classical gravity.
This is an important research direction due to the foreseeable improvements of
gravitational wave observations over the next decade(s), demanding ever more accurate predictions
for the motion and radiation of compact astrophysical binaries.

An improvable part of our double-copy procedure is the split of worldline vertices
with more than one gluon, using a delta distribution of the worldline parameter.
This leads to singular terms involving $\delta(0)$ in the quantum corrections to the
effective action, which we boldly drop in our classical consideration. While one cannot expect a consistent treatment
of all quantum corrections in an approach based on classical worldlines, addressing this
problem might still lead to a crucial improvement of our approach. An interesting solution is
to integrate out the auxiliary field $\psi$ along the worldline. Since its propagator is a
step function, the delta distributions used to split worldline vertices can be produced
by derivatives in the numerators acting on these worldline propagators. This will avoid the terms
$\delta(0)$. However, after the double-copy step, this will produce an auxiliary field $\psi$
propagating along the worldline in the dilaton gravity, too. But it appears possible to
surgically remove this unwanted feature from the result. Alternatively, one can possibly give
the auxiliary field an additional property that one desires of the theory, like a spin.
It is interesting to note that step-function propagators along the worldline basically
introduce a time ordering. This time ordering is also present in the Wilson loop, which
can be used to calculate the potential between a quark-antiquark pair.

We may also compare our work to a direct calculation of radiation via equations of motion in
\cite{Shen:2018ebu,Goldberger:2016iau,Goldberger:2017vcg,Goldberger:2017ogt}, which is
based on the same classical color charges (Yang-Mills) and point-masses (dilaton gravity) as our work.
If we descend from the effective action to the equations of motion through a variation, then derivatives
of the propagators may appear. These derivatives are not part of the numerators which take part
in the double copy. Adapting the double copy at the level of equations of motion
then faces the challenge of identifying numerators and (derivatives of) propagators.
This problem was elegantly solved by Shen \cite{Shen:2018ebu}:
replacing the Yang-Mills numerators by a copy of the color factor should lead to the corresponding
result in bi-adjoint scalar theory, thus providing a prescription to separate
numerator and propagator structures. Here we avoided this complication by working at the
action level. On the other hand, it is straightforward to calculate the (gauge independent)
radiation emitted by a binary system when working with equations of motion
\cite{Shen:2018ebu,Goldberger:2016iau,Goldberger:2017vcg, Goldberger:2017ogt}.

While maybe less straightforward, extending our work to radiation from binaries
will be an interesting future direction. Note that the calculation of classical binary
interaction potentials presented here was an effective field theory approach similar to
\cite{Goldberger:2004jt}. Following this line of research \cite{Goldberger:2009qd}, the
calculation of radiation aided by a double copy (at the level of the action) could succeed.
Past work also suggests that an extension of our work to spinning point particles
\cite{Goldberger:2017ogt} or a projection to pure general relativity \cite{Johansson:2014zca,
Luna:2017dtq} will be possible. But the most important future extension of our present paper will be
to demonstrate the double-copy prescription at higher orders. At next-to-next-to-leading order
in the effective action,
the BCJ color-kinematics duality (Jacobi identity) will start to play a crucial role in identifying
suitable color factors and numerators for the double copy. As for all work in the area of
color-kinematics duality and double copy, it will be highly interesting to see if our proposal
(or a modification thereof) survives at the next order. For the original BCJ case, the answer
to this was so far always positive.

\acknowledgments
We wish to thank Alessandra Buonanno, Thibault Damour, Radu Roiban, Chia-Hsien Shen and
Justin Vines for helpful discussions.

\appendix

\section{Conventions and Feynman rules}\label{conventions}
The signature of spacetime is $-2$ (``mostly minus''). We use units such that the speed of
light $c=1$ and use $c^{-1}$ as a formal post-Newtonian counting parameter.

The Yang-Mills action is the standard one, $S_\text{YM}= -\frac{1}{4}\int d^{4}x\, (F_{\mu\nu}^{a})^{2}$, with
the color convention $\text{tr}(T^{a}T^{b}) = \delta^{ab} / 2$. 
The field strength $F_{\mu\nu}^a$ is given by
\begin{multline}
  [ D_\mu, D_\nu ] = - i g F_{\mu\nu}^a T^a \\
  \Rightarrow \, F_{\mu\nu}^a = \partial_\mu A^a_\nu - \partial_\nu A^a_\mu + g f^{bca} A^b_\mu A^c_\nu ,
\end{multline}
where $D_\mu = \partial_\mu - i g A^a_\mu T^a$, $[ T^a, T^b ] = i f^{abc} T^c$ and the structure constants $f^{abc}$ are
totally antisymmetric here.
Working in Feynman gauge we have the coordinate space Feynman rules
of Yang-Mills theory,
	\begin{align}
	\langle A_{\mu}^{a}(x)\, A_{\nu}^{b}(y)\rangle_{0}& = \qquad
	\raisebox{-0.3 cm}{
		\begin{fmfgraph*}(20,20)
		\fmfpen{thick}
		\fmfleft{s1}
		\fmfright{s2}
		\fmf{photon,width=1}{s1,s2}
		\fmfv{label=$b$ $\nu$,label.dist=2.5}{s2}
		\fmfv{label=$a$ $\mu$,label.dist=2.5}{s1}
		\end{fmfgraph*}} 
	\qquad    =  \frac{\hbar}{i}\, \eta_{\mu\nu}\, \delta_{ab}\, D(x-y) , \\
	\raisebox{-0.6cm}{
		\begin{fmfgraph*}(40,40)
		\fmfpen{thick}
		\fmfleft{s1,s2}
		\fmfright{h1}
		\fmf{vanilla,width=1}{s1,v1,s2}
		\fmffreeze
		\fmf{photon,width=1}{h1,v1}
		\fmfdot{v1}
		\fmfv{label=$\tau$,label.dist=2.5}{v1}
		\fmfv{label=$a$ $\mu$,label.dist=2.5}{h1}
		\end{fmfgraph*}}
	\qquad & = \frac{i}{\hbar}\, 2 g\, \lambda(\tau) \, p^{\mu}(\tau)\, c^{a}(\tau) , \\[2ex]
	\raisebox{-0.6cm}{
		\begin{fmfgraph*}(40,40)
		\fmfpen{thick}
		\fmfleft{s1,s2}
		\fmfright{h1,h2}
		\fmf{vanilla,width=1}{s1,v1,s2}
		\fmffreeze
		\fmf{photon,width=1}{h1,v1}
		\fmf{photon,width=1}{h2,v1}
		\fmfdot{v1}
		\fmfv{label=$\tau$,label.dist=2.5}{v1}
		\fmfv{label=$a$ $\mu$,label.dist=2.5}{h1}
		\fmfv{label=$b$ $\nu$,label.dist=2.5}{h2}
		\end{fmfgraph*}}
	\qquad & = \frac{i}{\hbar} 2 g^{2}\,\lambda(\tau)\, c^{a}(\tau)\, c^{b}(\tau)\, \eta^{\mu\nu}, \\[2ex]
	\raisebox{-0.4cm}{
		\begin{fmfgraph*}(30,30)
		\fmfpen{thick}
		\fmfsurround{h1,s1,s2}
		\fmf{photon,width=1}{h1,v1}
		\fmf{photon,width=1}{s1,v1}
		\fmf{photon,width=1}{v1,s2}
		\fmfdot{v1}
		\fmfv{label=$c$ $\mu_{3}$,label.dist=2.5}{h1}
		\fmfv{label=$b$ $\mu_{2}$,label.dist=2.5}{s2}
		\fmfv{label=$a$ $\mu_{1}$,label.dist=2.5}{s1}
		\end{fmfgraph*}}
	\qquad & =
                 - \frac{i}{\hbar} g\, f^{abc}\, V_{123}^{\mu_1\mu_2\mu_3} ,
	\end{align}\\
where we used the abbreviation $ V_{123}^{\mu_1\mu_2\mu_3}$ from \eqref{defV}.

Our definition for the Riemann tensor is $R^\mu{}_{\nu\alpha\beta} = \partial_\alpha\Gamma^\mu{}_{\nu\beta} - \partial_\beta \Gamma^\mu{}_{\nu\alpha}
+ \Gamma^\mu{}_{\lambda\alpha} \Gamma^\lambda{}_{\nu\beta}
- \Gamma^\mu{}_{\lambda\beta} \Gamma^\lambda{}_{\nu\alpha}$,
from which the Ricci tensor $R_{\mu\nu} = R^\alpha{}_{\mu\alpha\nu}$ and Ricci scalar
$R = R^\mu{}_\mu$ are obtained through contractions.
The Feynman rules for dilaton gravity, for our choice of gauge and field redefinitions,
follow straightforwardly from \eqref{Sdgexpand} and \eqref{eq:GRWL} as
	\begin{align}
          \langle g_{\mu\nu}(x)\, g_{\alpha\beta}(y)\rangle_{0} &= \qquad
	\raisebox{-0.3 cm}{
		\begin{fmfgraph*}(20,20)
		\fmfpen{thick}
		\fmfleft{s1}
		\fmfright{s2}
		\fmf{dbl_wiggly,width=1}{s1,s2}
		\fmfv{label=$\mu\nu$,label.dist=2.5}{s2}
		\fmfv{label=$\alpha\beta$,label.dist=2.5}{s1}
		\end{fmfgraph*}} 
	\qquad    = - \frac{\hbar}{i} P_{\mu\nu\alpha\beta} D(x - y) , \\
	\raisebox{-0.6cm}{
		\begin{fmfgraph*}(40,40)
		\fmfpen{thick}
		\fmfleft{s1,s2}
		\fmfright{h1}
		\fmf{vanilla,width=1}{s1,v1,s2}
		\fmffreeze
		\fmf{dbl_wiggly,width=1}{h1,v1}
		\fmfdot{v1}
		\fmfv{label=$\tau$,label.dist=2.5}{v1}
		\fmfv{label=$\mu\nu$,label.dist=2.5}{h1}
		\end{fmfgraph*}}
	\qquad & = - \frac{i}{\hbar} \, \kappa \, \lambda(\tau) \, p^\mu(\tau) \, p^\nu(\tau) , \\[2ex]
	\raisebox{-0.6cm}{
		\begin{fmfgraph*}(50,40)
		\fmfpen{thin}
		\fmfleft{a1,a2}
		\fmfright{v2,v6}
		\fmf{vanilla}{a1,v1,a2}
		\fmffreeze
		\fmf{dbl_wiggly}{v1,v2}
		\fmf{dbl_wiggly}{v1,v6}
		\fmfdot{v1}
		\fmfv{label=$\tau$,label.dist=2.5}{v1}
		\fmfv{label=$\mu\nu$,label.dist=2.5}{v2}
		\fmfv{label=$\alpha\beta$,label.dist=2.5}{v6}
		\end{fmfgraph*}}
          \qquad &= \frac{i}{\hbar}\, \kappa^2 \, \lambda(\tau) \, p^{(\mu}(\tau) \, \eta^{\nu)(\alpha} \, p^{\beta)}(\tau) , \label{twograv} \\[2ex]
	\raisebox{-0.4cm}{
		\begin{fmfgraph*}(30,30)
		\fmfpen{thick}
		\fmfsurround{h1,s1,s2}
		\fmf{dbl_wiggly,width=1}{h1,v1}
		\fmf{dbl_wiggly,width=1}{s1,v1}
		\fmf{dbl_wiggly,width=1}{v1,s2}
		\fmfdot{v1}
		\fmfv{label=$\mu_{3}\nu_{3}$,label.dist=2.5}{h1}
		\fmfv{label=$\mu_{2}\nu_{2}$,label.dist=2.5}{s2}
		\fmfv{label=$\mu_{1}\nu_{1}$,label.dist=2.5}{s1}
		\end{fmfgraph*}}
	\qquad & = \frac{i}{\hbar} \, \frac{\kappa}{4} \,
	V_{123}^{\alpha_1\alpha_2\alpha_3} \, V_{123}^{\beta_1\beta_2\beta_3} \prod_{i=1}^3 P_{\alpha_i\beta_i}{}^{\mu_i\nu_i} ,
	\end{align}\\
        where $P_{\mu\nu\alpha\beta} = \frac{1}{2} ( \eta^{\mu\alpha} \eta^{\nu\beta} + \eta^{\nu\alpha} \eta^{\mu\beta} )$.
        
By comparing the Yang-Mills and dilaton-gravity Feynman rules, it is clear that the double
copy of many diagrams is trivial, since the vertices already show the double-copy
structure, except for the vertex \eqref{twograv}.

\section{Self-interactions in the post-Newtonian approximation}\label{self}
In this appendix, we wish to discuss the self-interactions, which vanish using the post-Newtonian
expansion of the propagator \eqref{propPN}. Using the gauge choice $\tau = t$,
the simplest self-interaction at leading order reads
\begin{align}
	\raisebox{-0.6cm}{
		\begin{fmfgraph}(50,40)
		\fmfpen{thin}
		\fmfleft{a1,a2}
		\fmfright{b1,b2}
		\fmf{vanilla}{a1,v1,a2}
		\fmf{vanilla}{b1,v2,v4,v5,b2}
		\fmffreeze
		\fmf{dbl_wiggly,left=1}{v2,v5}
		\fmfdot{v2,v5}
              \end{fmfgraph}} &= - \frac{i \kappa^2}{2 \hbar} \int dt dt' \, [p(t) \cdot p(t')]^2 D(x(t) - x(t'))  \\[2ex]
\begin{split}
	&= - \frac{i \kappa^2}{2 \hbar} \int \frac{d^3\vct{k} dt dt'}{(2\pi)^3} [p(t) \cdot p(t')]^2
	\frac{e^{i \vct{k} \cdot [\vct{x}(t) - \vct{x}(t')]}}{\vct{k}^2} \nl
        \times \left[ 1 + \frac{\partial_t \partial_{t'}}{\vct{k}^2} + \dots \right] \delta(t-t') .
\end{split}
\end{align}
After performing one of the time integrations, the times are identified
$t=t'$ and in the end $e^{i \vct{k} \cdot [\vct{x}(t) - \vct{x}(t')]} = 1$.
The $\vct{k}$-integral then turns into a scaleless one and vanishes in dimensional regularization.
Hence one gets a vanishing result at each post-Newtonian order. This result is physically sensible.
Indeed, a crucial property of the propagator (in Minkowski signature) is that its support
is on the light-cone. Now, the worldlines of massive particles can not intersect the
light cones emanating from them, so that these self-interaction
should be zero. Similar arguments apply to the other self-interactions at next-to-leading order
in the effective action
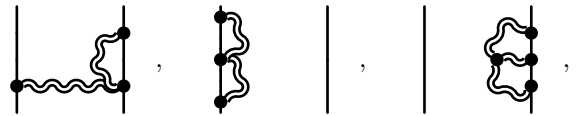
\begin{center}
\raisebox{-0.6cm}{
		\begin{fmfgraph}(50,40)
		\fmfpen{thin}
		\fmfleft{a1,a2}
		\fmfright{b1,b2}
		\fmf{vanilla}{a1,v1,a,s,a2}
		\fmf{vanilla}{b1,v2,v4,v5,b2}
		\fmffreeze
		\fmf{dbl_wiggly}{v1,v2}
		\fmf{dbl_wiggly,left=1}{v2,v5}
		\fmfdot{v1,v2,v5}
		\fmfv{label=$\tau_2$}{v1}
		\fmfv{label=$\ttau_3$}{v5}
		\fmfv{label=$\ttau_1$}{v2}
		\end{fmfgraph}} , \quad
\raisebox{-0.6cm}{
		\begin{fmfgraph}(50,40)
		\fmfpen{thin}
		\fmfleft{a1,a2}
		\fmfright{b1,b2}
		\fmf{vanilla}{b1,v1,b2}
		\fmf{vanilla}{a1,v2,g,h,j,v4,a,s,d,v5,a2}
		\fmffreeze
		\fmf{dbl_wiggly,right=1,tension=0}{v2,v4}
		\fmf{dbl_wiggly,right=1,tension=0}{v4,v5}
		\fmfdot{v2,v4,v5}
		\fmfv{label=$\tau_1$}{v4}
		\fmfv{label=$\tau_2$,l.a=180}{v2}
		\fmfv{label=$\tau_3$,l.a=180}{v5}
		\end{fmfgraph}} , \quad
\raisebox{-0.6cm}{
			\begin{fmfgraph}(50,40)
				\fmfpen{thin}
				\fmfleft{a1,h1,a2}
				\fmfright{b1,b2}
				\fmf{vanilla}{a1,a2}
				\fmf{vanilla}{b1,v2,v1,v3,b2}
				\fmffreeze
				\fmf{dbl_wiggly,left=1}{v2,z1,v3}
				\fmf{dbl_wiggly}{z1,v1}
				\fmf{phantom,tension=2}{h1,z1,v1}
				\fmfdot{v1,z1,v2,v3}
		\end{fmfgraph}} ,
\end{center}	
which all vanish in dimensional regularization.

\end{fmffile}%


%

\end{document}